\documentclass[12pt]{article}

\usepackage{scicite}

\usepackage{graphicx}
\usepackage{hyperref}
\usepackage{times}

\newenvironment{sciabstract}{%
\begin{quote} \bf}
{\end{quote}}

\newcounter{lastnote}

\title{Spin-mixing enhanced proximity effect in aluminum-based superconductor-semiconductor hybrids}

\author
{G.P. Mazur$^{1\ast\dagger}$, N. van Loo$^{1\dagger}$, J.Y. Wang$^{1}$, T. Dvir$^{1}$,  \\
G. Wang$^{1}$, A. Khindanov$^{2}$, S. Korneychuk$^{1}$, F. Borsoi$^{1}$,\\
 R.C. Dekker$^{1}$,  G. Badawy$^{3}$, P. Vinke$^{1}$, S. Gazibegovic$^{3}$,\\ E.P.A.M. Bakkers$^{3}$, M. Quintero-P\'erez$^{1,4}$, S. Heedt$^{1}$, L.P. Kouwenhoven$^{1}$\\
\normalsize{$^{1}$ QuTech and Kavli Institute of Nanoscience,}  \\
\normalsize{Delft University of Technology, 2600 GA Delft, The Netherlands}\\
\normalsize{$^{2}$ Department of Physics, University of California, Santa Barbara,} \\
\normalsize{California 93106, USA}\\
\normalsize{$^{3}$ Department of Applied Physics, Eindhoven University of Technology,} \\ 
\normalsize{5600 MB Eindhoven, The Netherlands}\\
\normalsize{$^{4}$ Netherlands Organisation for Applied Scientific Research (TNO), Delft, The Netherlands,} \\ 
\normalsize{2600 AD, Delft, The Netherlands}\\
\normalsize{$^\ast$To whom correspondence should be addressed. E-mail:  g.p.mazur@tudelft.nl}\\
\normalsize{$^\dagger$ These authors contributed equally to this work.}
}

\date{}

\begin{document} 


\baselineskip24pt


\maketitle

\newpage
\begin{sciabstract}
In superconducting quantum circuits, aluminum is one of the most widely used materials. It is currently also the superconductor of choice for the development of topological qubits. In this application, however, aluminum-based devices suffer from poor magnetic field compatibility. In this article, we resolve this limitation by showing that adatoms of heavy elements (e.g.\ platinum) increase the critical field of thin aluminum films by more than a factor of two. Using tunnel junctions, we show that the increased field resilience originates from spin-orbit scattering introduced by Pt. We exploit this property in the context of the superconducting proximity effect in semiconductor-superconductor hybrids, where we show that InSb nanowires strongly coupled to Al/Pt films can maintain superconductivity up to 7\,T. The two-electron charging effect, a fundamental requirement for topological quantum computation, is shown to be robust against the presence of heavy adatoms. Additionally, we use non-local spectroscopy in a three-terminal geometry to probe the bulk of hybrid devices, showing that it remains free of sub-gap states. Finally, we demonstrate that semiconductor states which are proximitized by Al/Pt films maintain their ability to Zeeman-split in an applied magnetic field. Combined with the chemical stability and well-known fabrication routes of aluminum, Al/Pt emerges as the natural successor to Al-based systems and is a compelling alternative to other superconductors, whenever high-field resilience is required.
\end{sciabstract}

\newpage
\section*{Introduction}
Topological superconductivity can arise in hybrid material stacks containing a conventional superconductor and a semiconductor with strong Rashba spin-orbit coupling~\cite{Lutchyn_2010:PRL,Oreg_2010:PRL}.  Narrow-gap semiconductors with a large g-factor and low carrier density (such as InAs and InSb) are most commonly used, either as 1D nanowires~\cite{krogstrup_2015:nmat} or 2D electron gases~\cite{shabani_2016:PRB}. On the superconductor side, aluminum has become the material of choice. Thin shells made of this metal combined with an oxide-free interface result in clean electronic transport~\cite{Heedt:2021_NC,Borsoi:2021_AFM,gazibegovic_2017:nature}. This includes suppressed sub-gap tunneling conductance (hard induced gap) and parity-conserving transport~\cite{shen_2018:ncomms}, enabling the search for topological superconductivity. For a topological phase to emerge, the minimal condition states that the Zeeman energy $V_{\rm Z} = g \mu_{\rm B} B$ must be larger than the induced superconducting gap $\Delta$, where $g$ is the Land\'e g-factor, $\mu_{\rm B}$ is the Bohr magneton and $B$ is the applied magnetic field. It was demonstrated recently that the properties of the semiconductor, such as spin-orbit coupling and g-factor, are renormalized by the presence of a proximitizing metal~\cite{Antipov:2018_PRX,de_Moor_2018:NJP}. As a result, stronger magnetic fields than initially anticipated are required to close and reopen the induced superconducting gap~\cite{Ahn_2021:PRM}. Typical aluminum-based hybrids, however, have a zero-field superconducting gap $\Delta_{0}$ ranging from 200 to 300\,$\mu$eV~\cite{lutchyn:2018_NRP}, which results in poor field compatibility. This has fueled the search for alternative superconducting systems, with recent works reporting superconductivity and parity-conserving transport in InSb/Sn~\cite{Pendharkar_2021:S}, InAs/Pb~\cite{Paajaste_NL_2015,Kanne:2021_NNano} and InAs/In\cite{Bjergerfelt_2021:NL} hybrids. These superconductors offer higher field compatibility than aluminum, yet they bring different challenges such as chemical instability and fabrication constraints~\cite{Baker_IBM:1980,Khan:2020_ACSNANO}. For a Bardeen-Cooper-Schrieffer (BCS) type superconductor, there are two dominant mechanisms which quench superconductivity in a finite magnetic field. The first of these is orbital depairing, which results from the cyclotron motion of conduction electrons due to the magnetic field. 
\newpage
\noindent
If the superconductor is grown as a thin film, this mechanism can be suppressed when the field is applied in the plane of the film. For light elements like Al, there is a second contribution arising from spin physics. Once the magnetic field reaches a certain value, the paramagnetic ground state becomes energetically favourable. This results in a first-order phase transition into the normal state. The field for which this happens is known as the Chandrasekhar-Clogston~\cite{Chandrasekhar1962,Clogston1962} or Pauli limit, and is given by $B_{P}=\Delta_{0}/(\sqrt{2}\,\mu_{B})$. In addition, the quasiparticle excitation spectrum spin-splits upon applying a magnetic field. In their seminal experiment~\cite{Tedrow_1982:PRB}, Tedrow and Meservey demonstrate that Zeeman splitting can be quenched and eventually suppressed completely through the addition of heavy metal impurities, such as platinum (Pt). These heavy atoms introduce spin-orbit scattering, which prevents spins from being polarized by an external magnetic field. As a result, superconductors made of lightweight elements can reach unprecedentedly high critical fields. This has straightforward applications in the field of semiconductor-superconductor hybrids, where large Zeeman energies are a necessary condition for achieving a topological superconducting phase.

\section*{Al/Pt thin films and tunnel junctions}
\begin{figure}[h!]
    \centering
	    \includegraphics[width=\textwidth]{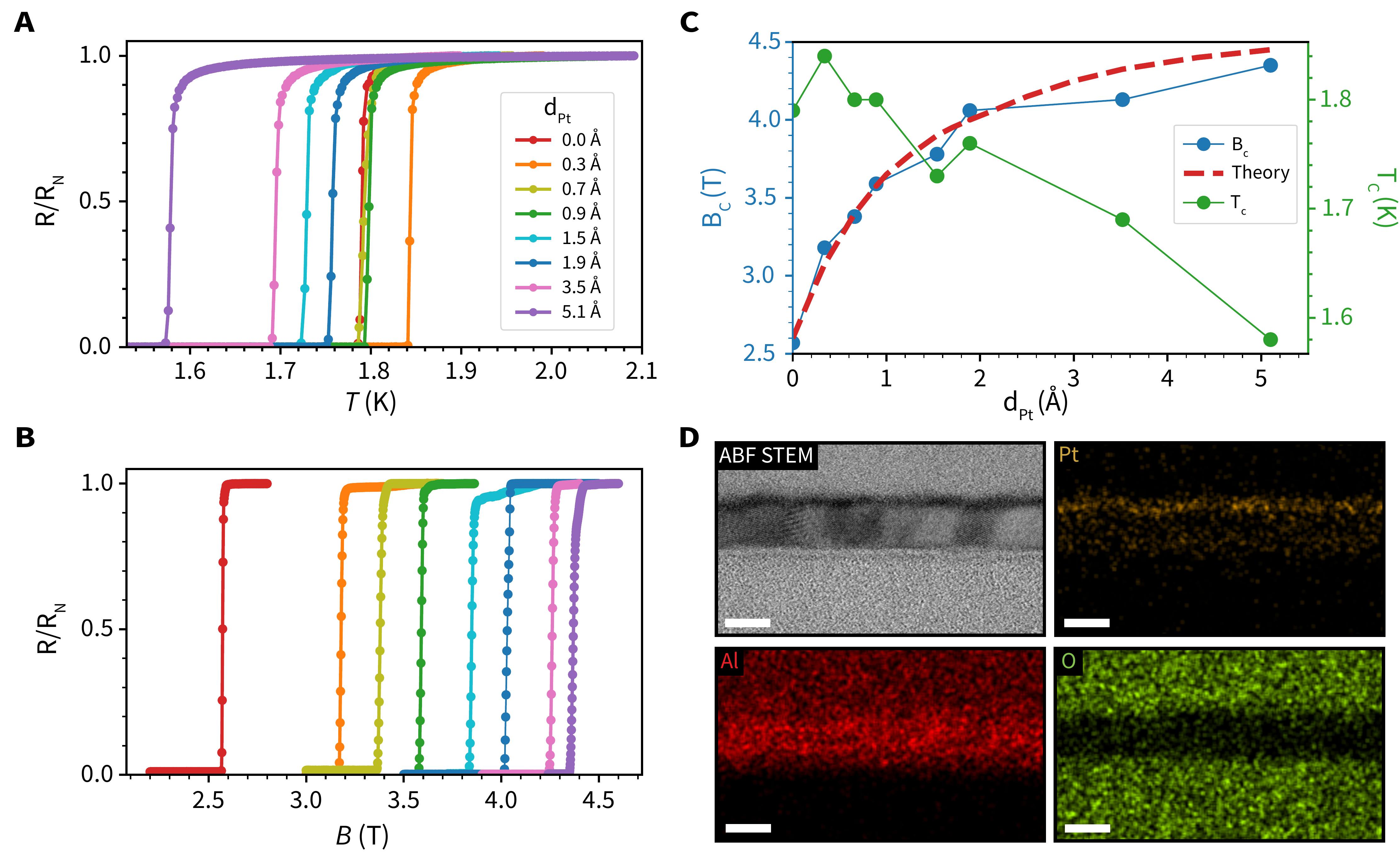}
	\caption{\textbf{Properties of Al/Pt thin films.} Four-point measurements of resistance $R$ normalized to its value in the normal state $R_{\rm N}$ as a function of temperature (\textbf{A}) and magnetic field (\textbf{B}). Measurements have been performed for 6\,nm thick aluminum films with varying amount of platinum $d_{\rm Pt}$. (\textbf{C}) Critical magnetic field and temperature as a function of Pt thickness, together with the predicted critical field from theory calculations. (\textbf{D}) Annular bright field scanning-tunneling electron micrograph and energy-dispersive X-ray images of the Al film with 1.89\,$\textrm{\AA}$ of Pt. Scale bars are 5\,nm.}
    \label{fig:Fig_1}
\end{figure}

We begin this study by evaluating the properties of aluminum films with a thickness of  6\,nm, coated with varying amounts of platinum. We define the platinum thickness $d_{\rm Pt}$ as measured by the quartz balance in the deposition chamber. Figure~\ref{fig:Fig_1} presents the superconducting transitions of Al/Pt films as a function of temperature (Fig.~\ref{fig:Fig_1}A) and parallel magnetic field (Fig.~\ref{fig:Fig_1}B). Importantly, the addition of Pt does not affect the shape and sharpness of the superconducting transitions, which indicates that the films do not become strongly disordered or inhomogeneous~\cite{Monceu:1974_PLA}. The bare aluminum film has a critical temperature $T_{\rm c}$ = 1.79\,K and a critical field $B_{\rm c}$ = 2.6\,T. Upon the addition of platinum, the critical field is increased above the bare aluminum's Chandrasekhar-Clogston limit already for $d_{\rm Pt} \approx 1$\,$\textrm{\AA}$, while leaving $T_{\rm c}$ unaffected. 
\\[2ex]
\noindent
In agreement with previous studies on Al/Pt bilayers~\cite{Tedrow_1982:PRB}, the critical field starts to saturate for $d_{\rm Pt} \approx 2$\,$\textrm{\AA}$ and increases only by an additional 300\,mT for $d_{\rm Pt} \approx 5.1$\,$\textrm{\AA}$ (see Fig.~\ref{fig:Fig_1}C). At these thicknesses, however, $T_{\rm c}$ starts to decrease as a result of the inverse proximity effect, as shown for Au/Be bilayers~\cite{Wu_2006:PRL}. Our theoretical model based on the Usadel equation (see supplementary materials~\cite{Supplement}) captures the increase of $B_{\rm c}$ as a direct result of including spin-orbit scattering (Fig.~\ref{fig:Fig_1}C). 
\newpage
The calculated spin-orbit scattering rates increase linearly with $d_{\rm Pt}$~\cite{Supplement}, in agreement with previous experiments~\cite{Tedrow_1982:PRB}. In addition, we perform a structural analysis of the films. Fig.~\ref{fig:Fig_1}D presents the cross-section of an Al/Pt film with $d_{\rm Pt} \approx 1.9$\,$\textrm{\AA}$, which reveals the poly-crystalline structure of the Al. The results of electron energy loss spectroscopy (EELS)~\cite{Supplement} performed on the studied samples indicate that aluminum and platinum do not form an alloy, in agreement with the bulk phase diagram~\cite{mcalister_1986:BAPD}. 

\begin{figure}[h!]
    \centering
	    \includegraphics[width=\textwidth]{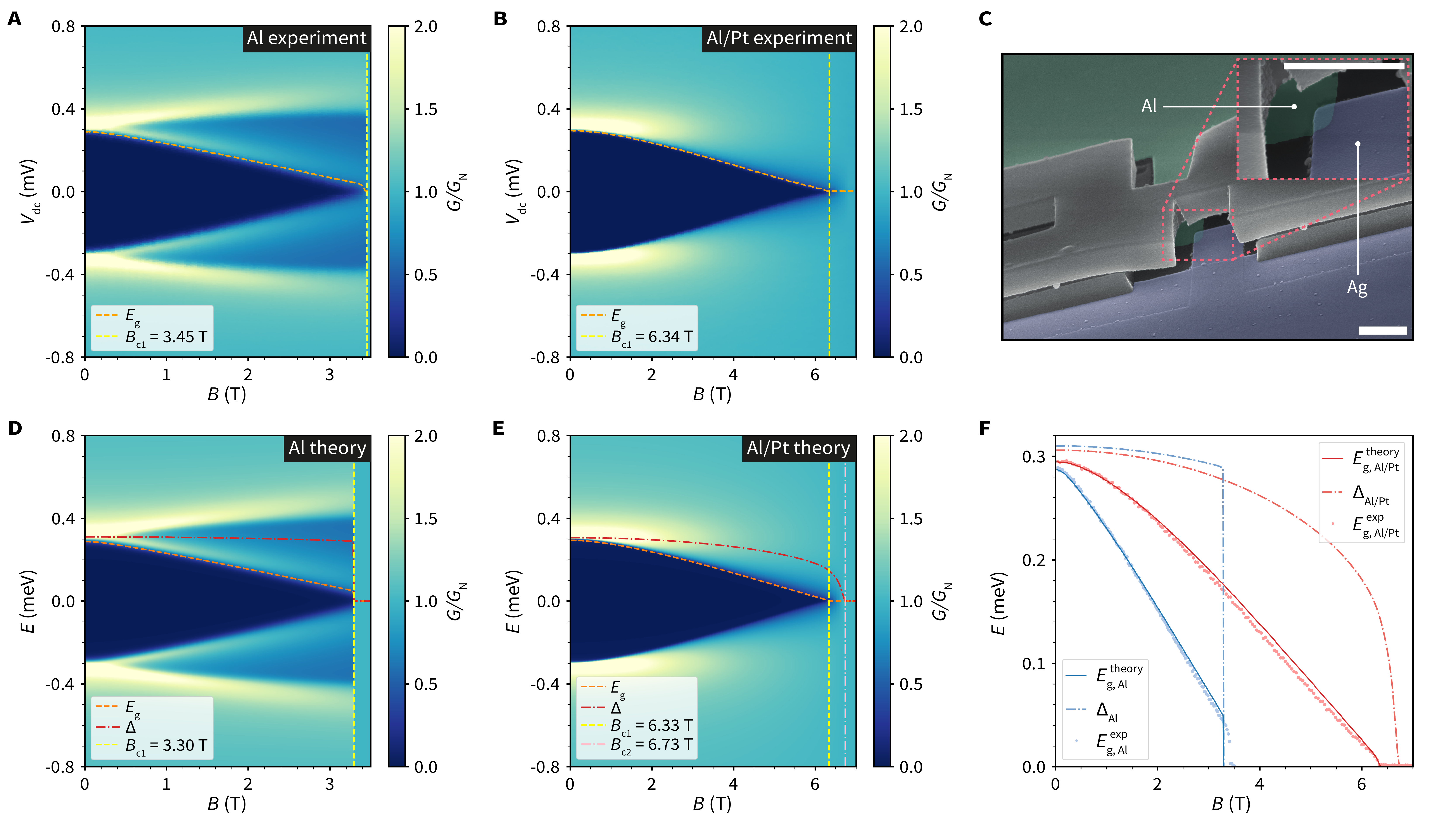}
	\caption{\textbf{Conductance spectroscopy on Al and Al/Pt NIS tunnel junctions.} (\textbf{A}) Experimental tunneling conductance of a $\sim$\,4.5\,nm Al tunnel junction. (\textbf{B}) Experimental tunneling conductance of a $\sim$\,4.5\,nm Al + 1.9\,$\textrm{\AA}$ Pt tunnel junction. (\textbf{C}) False-color scanning-electron micrograph of a typical Al/AlOx/Ag tunnel junction. Scale bars are 1\,$\mu$m. (\textbf{D}) Tunneling conductance from theory calculations of the Al tunnel junction. (\textbf{E}) Tunneling conductance from theory calculations of the Al/Pt tunnel junction. The dashed orange lines present the energy gap $E_{\rm g}$. The order parameter $\Delta$ which is extracted from theory is presented by dashed red lines. Dashed yellow lines show the magnetic field $B_{\rm c1}$ for which the energy gap is closed, and the dashed  pink lines indicate the magnetic field $B_{\rm c2}$ for which the order parameter is calculated to vanish. (\textbf{F}) Overview of the extracted energy gap from experiments, the predicted energy gap from theory and the corresponding order parameter of the films.}
    \label{fig:Fig_2}
\end{figure}

Furthermore, we investigate the impact of Pt atoms on the Al quasiparticle density of states through normal-metal/insulator/superconductor (NIS) tunneling measurements. A typical device used for these measurements is shown in Figure~\ref{fig:Fig_2}C. For the aluminum film, we observe Zeeman splitting of the quasiparticle coherence peaks (Fig.~\ref{fig:Fig_2}A). At a magnetic field of $B$ = 3.45\,T, the film undergoes a first-order phase transition to the normal state. Our theoretical model reproduces these two key features (Fig.~\ref{fig:Fig_2}D), where the first-order transition is reflected in an abrupt collapse of the order parameter. The critical field extracted from the model is 200\,mT smaller than the experimentally measured value. This discrepancy between theory and experiment can be explained by the hysteretic behavior of the order parameter near the transition~\cite{Wu_PRB:1995} (see supplementary materials~\cite{Supplement}). A metastable superconducting state can persist for magnetic fields slightly above the calculated critical value.
For the Al/Pt film with $d_{\rm Pt}$ = 1.9\,\AA, Zeeman splitting is not observed. Instead, the film undergoes a second-order phase transition at $B$ = 6.34\,T induced primarily by orbital effects (Fig.~\ref{fig:Fig_2}B). Importantly, the energy gap in the film remains free of quasiparticle states. Theoretical modelling of the film reveals a small magnetic field range with gapless superconductivity close to the transition (Fig.~\ref{fig:Fig_2}E), which is an expected feature when the transition from the superconducting into the normal state is of second order\cite{Tinkham2004}. For both Al and Al/Pt films, the model yields diffusion constants which correspond to a mean free path of $l_{\rm mfp}\approx$ 7\,\AA. 
\newpage
This value is consistent with reports on Al films grown under similar conditions~\cite{Alegria_NN:2021,Meservey_1994:PR}. Since the addition of Pt does not seem to affect the mean free path, the increase in critical magnetic field cannot be attributed to increased disorder. The suppression of Zeeman splitting instead demonstrates that spin mixing is the dominant mechanism. From the model, the increased spin-orbit scattering rate of the Al/Pt film is extracted to be $\Gamma_{\rm SO}$ = 7.5\,meV, corresponding to a spin-orbit scattering time of $\tau_{\rm SO}=1.3\cdot10^{-13}$\,s. We note, however, that this extracted value of the spin-orbit scattering rate could be overestimated due to the presence of Fermi-liquid effects~\cite{Alexander_PRB:1985} (see discussion in supplementary materials~\cite{Supplement}). In Fig.~\ref{fig:Fig_2}F, the measured energy gap is shown together with the energy gap extracted from theory, as well as the corresponding order parameter. We observe good quantitative agreement between the model and our experiment.

\newpage
\section*{Spectroscopy and Coulomb blockade of InSb/Al/Pt hybrids}
The next step of our study is to induce superconductivity in InSb nanowires using Al/Pt films. In order for any material combination to be considered for Majorana experiments and topological qubits, two fundamental properties need to be demonstrated. In tunneling spectroscopy, an proximity-induced gap free of sub-gap states (i.e.\ a hard gap) should be observed. 
\\[2ex]
While this is conventionally done on hybrids with a grounded superconductor, designs of topological qubits typically contain superconducting segments which are floating~\cite{Plugge:2017_NJP}. These have a finite charging energy, and it is energetically favorable to add charges to such an island in pairs if the low-energy excitation spectrum of the hybrid is free of single-charge states (i.e.\ 2\textit{e} charging). Both a hard superconducting gap and 2\textit{e} charging have already been demonstrated for Al-based hybrids~\cite{Albrecht:2016_N,shen_2018:ncomms,krogstrup_2015:nmat,gazibegovic_2017:nature}. In order to confirm that platinum does not compromise these properties, for example through hosting single-electron states~\cite{Savin:2007_APL}, hybrids with a grounded superconducting shell as well as with a floating shell have been investigated. 

\begin{figure}[h!]
    \centering
        \includegraphics[width=\textwidth]{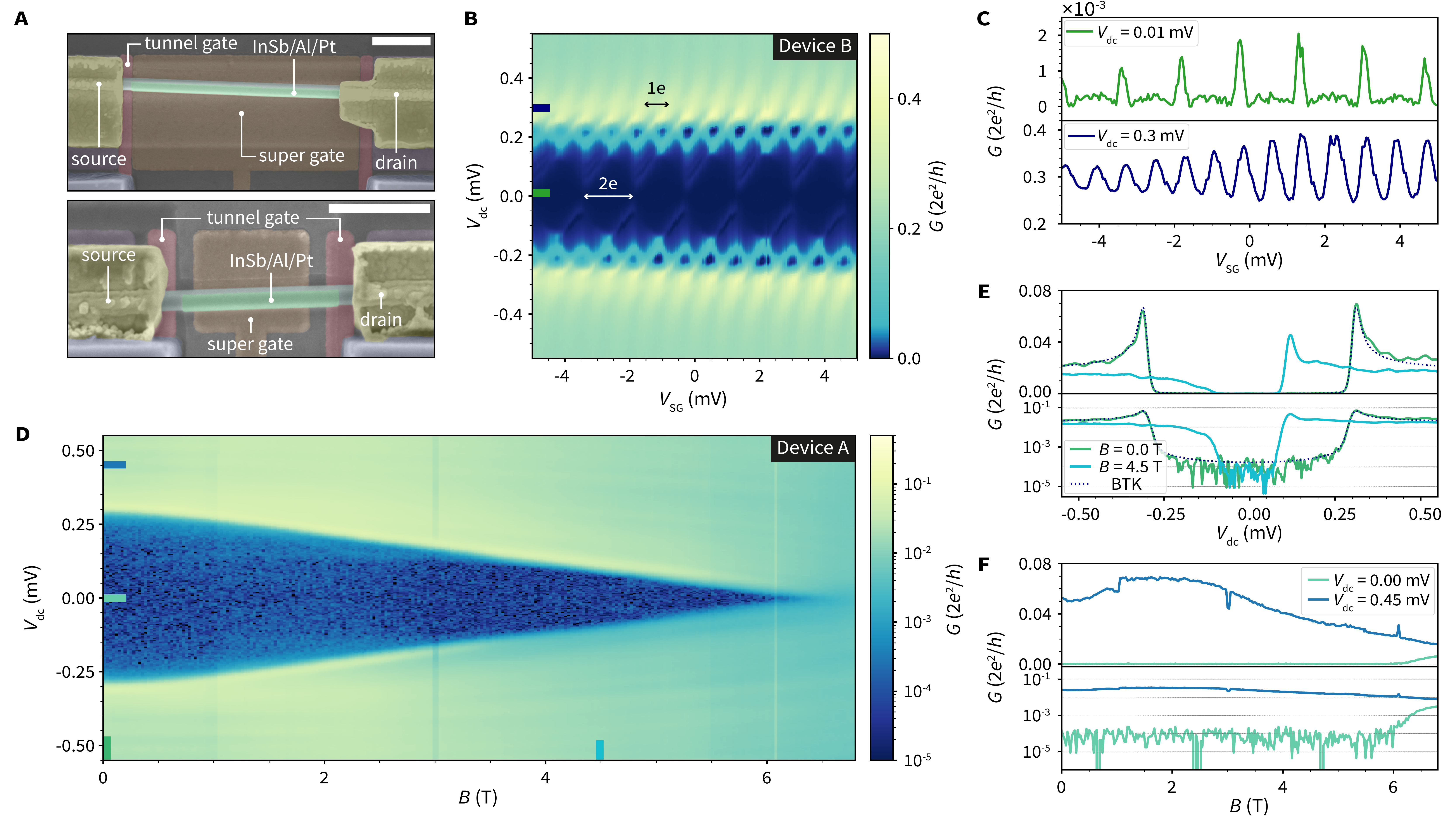}
	\caption{\textbf{Transport data on 2-terminal InSb/Al/Pt hybrids (devices A and B)}. (\textbf{A}) Examples of false-color scanning electron micrographs. Scale bars are 1\,$\mu$m. \textbf{Top}: Tunneling spectroscopy device with 1.8\,$\mu$m grounded superconducting segment. \textbf{Bottom}: Superconducting island device with 0.8\,$\mu$m floating superconducting segment.  (\textbf{B})  2$e$-periodic Coulomb diamonds measured on device B. (\textbf{C}) Linecuts from the Coulomb-blockade measurements at low bias (top) and finite bias (bottom). (\textbf{D}) Differential conductance from tunneling spectroscopy of device A in logarithmic scale, taken at $V_{\rm SG}$ = -1\,V as a function of parallel magnetic field. (\textbf{E}) Linecuts from the tunneling spectroscopy measurement, shown in linear (top) and logarithmic (bottom) scale. The dashed lines show conductance from BTK theory, with $\Delta$ = 304\,$\mu$eV, temperature $T$ = 70\,mK and transmission $G_{\rm N}$ = 0.018\,$\rm G_0$. (\textbf{F}) Differential conductance taken from the tunneling spectroscopy measurement at zero and finite bias, shown in linear (top) and logarithmic (bottom) scale.}
    \label{fig:Fig_3}
\end{figure}

In Figure~\ref{fig:Fig_3}, we show the results of both tunneling spectroscopy and Coulomb blockade measurements on InSb/Al/Pt nanowires. The fabrication follows our shadow-wall lithography method~\cite{Heedt:2021_NC,Borsoi:2021_AFM}, of which details can be found in the supplementary materials~\cite{Supplement}. In Fig.~\ref{fig:Fig_3}A, examples of a tunneling spectroscopy device (top) and a superconducting island device (bottom) are shown. The measurements are conducted by applying a bias voltage between the source and drain contacts. The chemical potential in the hybrid is controlled by the so-called super gate voltage $V_{\rm SG}$, while the tunnel gate voltages $V_{\rm TG}$ determine the density in the nanowire junctions.

For the spectroscopy device, the junction is tuned into the tunneling regime. Under this condition, the measured differential conductance reflects the quasiparticle density of states (DOS) in the proximitized section of the nanowire. Here, the super gate voltage is set to $V_{\rm SG}$ = -1\,V, where a strong coupling between the nanowire and the superconducting shell is expected~\cite{de_Moor_2018:NJP,Shen_2021:PRB}.
The differential conductance is shown as a function of magnetic field parallel to the nanowire axis in Fig.~\ref{fig:Fig_3}D, with linecuts taken at $B$ = 0.0\,T and $B$ = 4.5\,T presented in Fig.~\ref{fig:Fig_3}E. At zero magnetic field, a large superconducting gap of $\Delta$ = 304\,$\mu$eV is observed. This is significantly larger than in the case of conventional Al-based hybrids, which is a direct consequence of the reduced thickness ($\sim4.5$\,nm) of the Al shell~\cite{Tedrow_1982:PRB}. 
\\[2ex]
In addition, the in-gap conductance is suppressed by two orders of magnitude and the differential conductance matches the BTK theory~\cite{Blonder:1982_PRB}, indicating that the superconducting gap is free of sub-gap states (i.e.\ a hard gap). Importantly, at B = 4.5\,T the superconducting gap is still on the order of $\sim 100$\,$\mu$eV, which allows to look for Majorana signatures at Zeeman energies which were not accessible before. Remarkably, as can be seen from the in-gap and out-of-gap linecuts in Fig.~\ref{fig:Fig_3}F, the superconducting gap remains hard up to $B$ = 6.0\,T. The field compatibility offered by Al/Pt hybrids opens up the opportunity to study high-field signatures of Majorana zero modes, like Majorana oscillations~\cite{DasSarma:2012_PRB}.


The superconducting island device is studied by inducing tunnel barriers in the nanowire junctions,  which separate the island from the leads. The voltage on the super gate is then swept to tune the charge on the island, as shown in Fig.~\ref{fig:Fig_3}B. This results in a periodic sequence of 2\textit{e} Coulomb diamonds with a charging energy $E_{\rm c}\approx$ 30\,$\mu$eV. Linecuts are shown in Fig.~\ref{fig:Fig_3}C, where at finite bias the Coulomb peak periodicity has doubled due to the onset of single-electron transport in the quasiparticle excitation spectrum. The magnetic field evolution is shown in the supplementary materials~\cite{Supplement}. The observation of 2\textit{e} charging demonstrates that semiconductors coupled to Al/Pt are a suitable replacement of Al-based hybrids, capable to be used for the development of parity-protected topological qubits.

\begin{figure}[h!]
    \centering
        \includegraphics[width=0.66\textwidth]{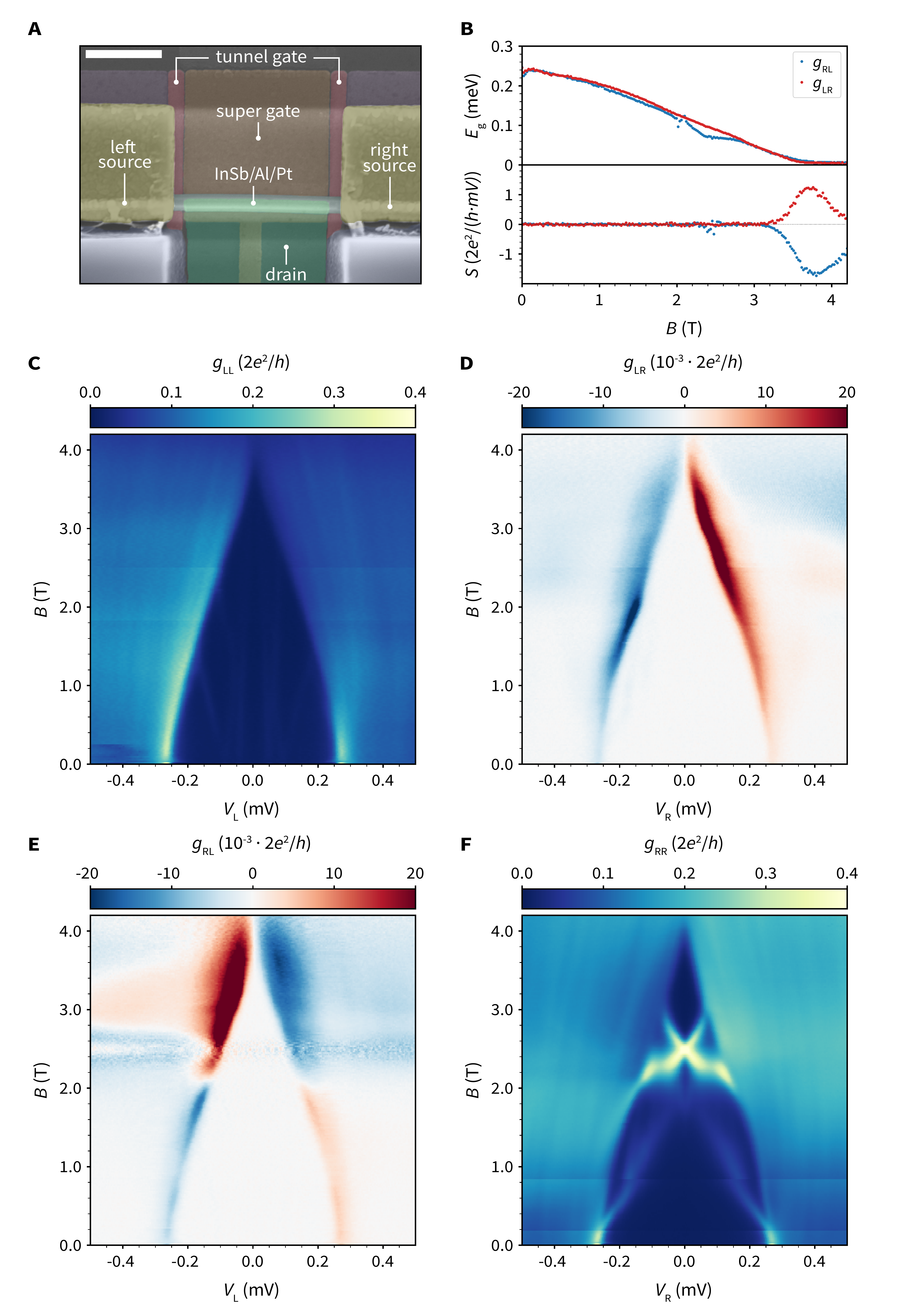}
	\caption{\textbf{Three-terminal measurements on InSb/Al/Pt hybrids (device C).} (\textbf{A}) Example of a false-color scanning electron micrograph, showing an InSb nanowire with 1\,$\mu$m hybrid segment. Scale bar is 500\,nm. (\textbf{B}) Extracted energy gap in the bulk of the hybrid (top), together with the non-local slope at zero bias (bottom). (\textbf{C-F}) Differential-conductance matrix measurements as a function of magnetic field. Panels \textbf{C,F} show the local conductances $g_{\rm LL}$ and $g_{\rm RR}$, respectively, whereas panels \textbf{D,E} illustrate the non-local conductances $g_{\rm LR}$ and $g_{\rm RL}$. Conductances are defined as $g_{\rm ij}$ $\equiv$ d$I_{\rm i}/$d$V_{\rm j}$. Data is taken at $V_{\rm SG}$ = -2\,V.} 
    \label{fig:Fig_4}
\end{figure}

\section*{Non-local measurements of three-terminal hybrids}
The experiments presented above are a prerequisite for investigating topological superconductivity. Most of the research to date has focused on the study of zero-bias anomalies  and their evolution as a function of chemical potential and magnetic field~\cite{lutchyn:2018_NRP,Prada:2020_NRP}. However, it is becoming increasingly clear that spectroscopy of the density of states at the ends of a nanowire is inconclusive when it comes to identifying an extended topological superconducting phase~\cite{Pan:2021_PRB,Puglia:2021_PRB}.
\\[2ex]
Instead, the simultaneous occurrence of zero-bias anomalies at both ends of a nanowire should be accompanied by the closing and reopening of the induced superconducting gap~\cite{pikulin:2021_arxiv}. While conventional tunneling experiments can provide information on the local density of states at both ends of a wire, the induced gap in the bulk of the hybrid can be probed by measuring the non-local conductance in a three-terminal geometry~\cite{Rosdahl:2018_PRB}. In Fig.~\ref{fig:Fig_4}A we present an SEM image of a typical device which is used for three-terminal measurements. 
\\[2ex]
The Al/Pt shell covers three facets of the InSb nanowire, and maintains a connection with the film on the substrate through which the device is connected to ground. The super gate voltage $V_{\rm SG}$ controls the chemical potential in the hybrid, and the tunnel gate voltages, $V_{\rm TL}$ and $V_{\rm TR}$, are used to control the left and right junction conductances, respectively.
In this work, we fix the super gate voltage to be $V_{\rm SG}$ = -2\,V, where the nanowire is expected to be strongly coupled to the superconducting shell~\cite{de_Moor_2018:NJP,Shen_2021:PRB}. Bias voltages are applied to the left ($V_{\rm L}$) or right ($V_{\rm R}$) normal contact while keeping the middle lead grounded. The local ($g_{\rm LL}$, $g_{\rm RR}$) and non-local ($g_{\rm LR}$, $g_{\rm RL}$) conductances are measured to form the full conductance matrix of the system, where they are defined as $g_{\rm ij}$ $\equiv$ d$I_{\rm i}/$d$V_{\rm j}$. Figure~\ref{fig:Fig_4}(C-F) shows an example of such a conductance matrix, measured as a function of parallel magnetic field. Due to the increased thickness of the shell ($\sim8$\,nm) compared with the single-facet device presented in the previous section, orbital contributions reduce the critical field of this hybrid. The local spectrum on the right junction exhibits a few sub-gap states, which are not present in the local spectrum on the left junction. This suggests that these states are confined locally near the right junction. The corresponding non-local conductances are zero everywhere inside the gap, confirming the local nature of these sub-gap states. This is emphasized in Fig.~\ref{fig:Fig_4}B, which displays the extracted energy gap $E_{\rm g}$ in the hybrid (top panel) as well as the normalized non-local slope $S$ (bottom panel). The non-local slope stays close to zero only while there is an energy gap present in the bulk of the hybrid. It starts to deviate from zero around $B_{\rm c}\approx$ 3.4\,T, indicating the gap in the system becomes soft before closing eventually around $B_{\rm c}\approx$ 3.8\,T. Remarkably, the induced superconducting gap in the bulk of these hybrids can be free of sub-gap states up to high magnetic fields. The effect of the super gate voltage on the proximity effect in these hybrids will be explored in an upcoming work.

\section*{Zeeman splitting inside the hybrid}
Having shown that the addition of Pt adatoms quenches the Zeeman effect in the Al shell, we turn our attention to the semiconductor part of the hybrid device. Breaking the Kramers degeneracy through Zeeman splitting of the DOS of the hybrid segment lies at the heart of the proposed schemes to reach the topological regime \cite{Lutchyn_2010:PRL,Oreg_2010:PRL}. 
\begin{figure}[h!]
    \centering
        \includegraphics[width=0.77\textwidth]{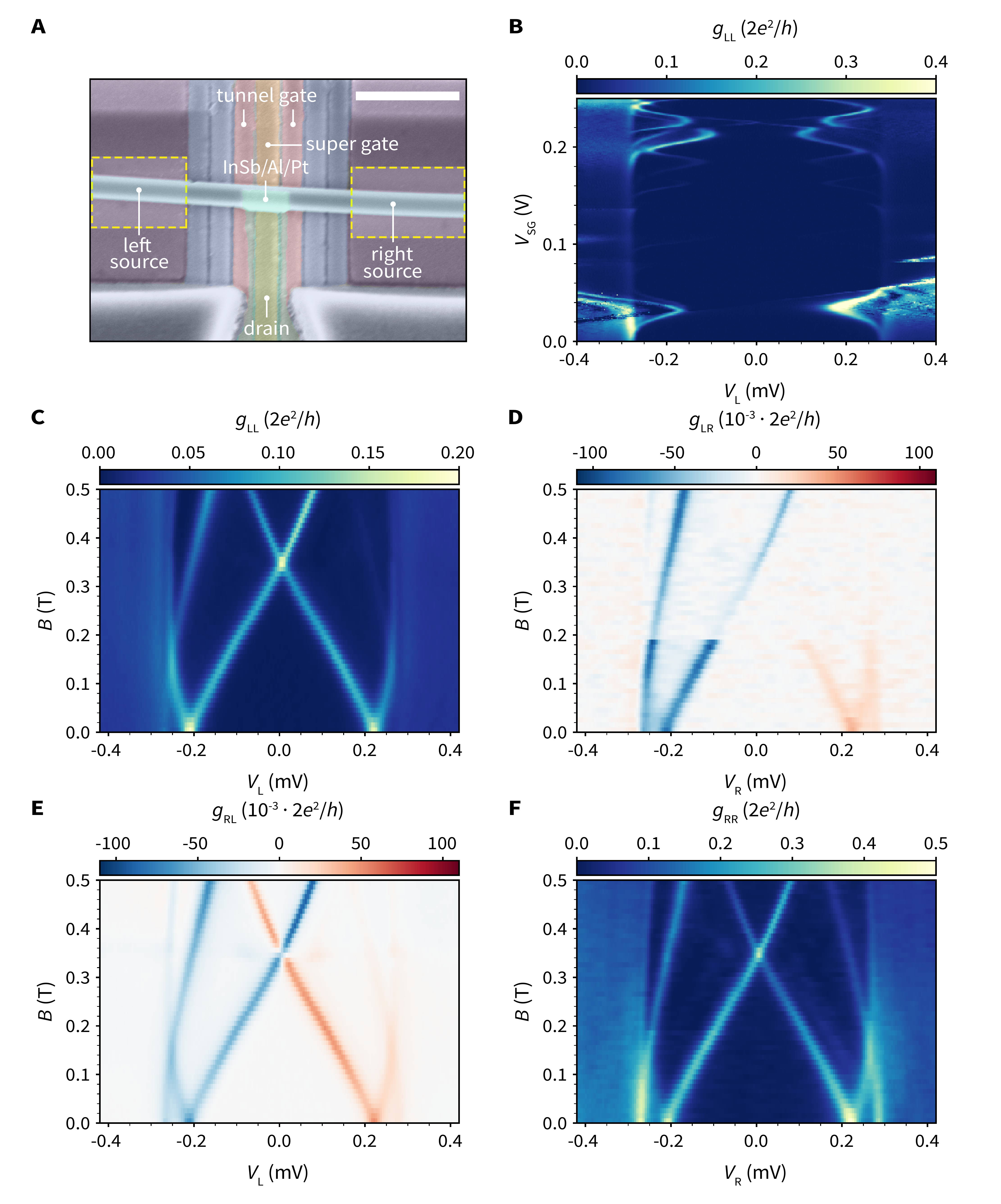}
	\caption{\textbf{Spin splitting of Andreev bound states in InSb/Al/Pt hybrids (device D).} (\textbf{A}) False-color scanning electron micrograph of the device, showing an InSb nanowire with 150\,nm hybrid segment. Scale bar is 500\,nm. Image is taken before contact deposition, and the contact design is indicated with yellow dashed rectangles. (\textbf{B}) Local differential conductance as a function of super gate voltage taken at zero applied magnetic field.  (\textbf{C-F}) Differential-conductance matrix measurements as a function of magnetic field. Panels \textbf{C,F} show the local conductances $g_{\rm LL}$ and $g_{\rm RR}$, respectively, whereas panels \textbf{D,E} illustrate the non-local conductances $g_{\rm LR}$ and $g_{\rm RL}$. Data is taken at $V_{\rm SG} = 0.0\,$V.} 
    \label{fig:Fig_5}
\end{figure}
Tunneling into discrete Andreev bound states (ABS) in the hybrid nanowire involves a transition between a spinless singlet state and spinful doublet states. The doublet state splits under the effect of an external magnetic field \cite{Lee_NN:2014, Jellingaard_PRB:2016}. Thus, measuring the evolution of the ABS spectrum in a magnetic field would show whether the effect of spin mixing leaks to the proximitized semiconductor.  
Figure~\ref{fig:Fig_5}A shows an SEM image of another three-terminal device, with a hybrid segment of 150\,nm. Here, each lead is separated from the hybrid by three finger gates. In this report, we apply a high voltage to the two finger gates (shown in blue) closest to each lead, effectively extending the normal leads. The tunnel gate closest to the hybrid segment is thus used to induce a tunnel barrier in the nanowire through which the hybrid segment is probed.  
In Fig.~\ref{fig:Fig_5}B we show the local conductance $g_{\rm LL}$ as a function of the $V_{\rm SG}$ and $V_{\rm L}$, taken at zero field. We observe a clean superconducting gap, in addition to a series of sub-gap resonances. They appear only when setting $V_{\rm SG} > 0\,$V, and reflect the presence of discrete states in the confined semiconductor. These states hybridize with the superconductor to form Andreev bound states. 
In Fig.~\ref{fig:Fig_5}C-F, we track the evolution of the ABS in an applied magnetic field by measuring the conductance matrix. We set $V_{\rm SG} = 0\,$V, so that the ABS energy is near its minimal value. To verify that the ABS is located in the hybrid segment and is not a local resonance on the left junction, we notice that it appears at the same energy on both sides, in  $g_{\rm LL}$ and $g_{\rm RR}$.  We also note that it shows up in the non-local signals, $g_{\rm RL}$ and $g_{\rm RL}$, consistent with a state which is extended along the entire hybrid. 
Upon application of the magnetic field, the ABS splits into two peaks that move with the same slope in opposite directions. 
\\[2ex]
The outgoing peaks are soon merged with the quasiparticle continuum, but the peaks shifting to lower energy cross at $B=0.34$\,T, where the ABS ground state turns from even to odd \cite{Lee_NN:2014}. We extract its gyromagnetic ratio to be $g=20.0\pm0.3$, showing only a moderate amount of renormalization of the semiconducting properties \cite{Antipov:2018_PRX, de_Moor_2018:NJP}.
Thus, the effect of spin mixing from Pt enhances the critical field of the Al shell, but does not negatively influence the spin properties in the semiconductor. This is evidenced by the picture of an extended ABS in the hybrid segment, which Zeeman splits with a high $g$ factor in the presence of a magnetic field. This demonstration is of crucial importance, as a spin-mixed hybrid would be fundamentally incapable of transitioning into a topological phase. It is still an open question if hybrids would preserve these properties when the semiconductors are coupled to high-atomic-number superconductors, like Sn,In or Pb \cite{Pendharkar_2021:S,Bjergerfelt_2021:NL,Kanne:2021_NNano}.

\section*{Conclusion}
In this work, we have examined the properties of thin aluminum films coated with sub-monolayer amounts of platinum, as well as semiconductor nanowires proximitized by Al/Pt bilayers. By measuring the critical temperature and magnetic field of thin films, we have found that $\sim$\,2\,$\textrm{\AA}$ of Pt can increase the critical field above the Chandrasekhar-Clogston limit, without having a significant effect on the size of the superconducting gap. We show, using our theoretical model, that the spin-orbit scattering rate of Pt-covered films is drastically increased. At the same time, various critical parameters of the films, such as the mean free path and coherence length, remain unaffected. When coupling InSb nanowires to these Pt-enhanced films, we observe a hard superconducting gap up to magnetic fields as high as 6\,T. Additionally, parity-conserving transport results in the formation of 2$e$-periodic diamonds in Coulomb-blockade experiments. Upon switching to a three-terminal geometry, non-local measurements provide evidence of a bulk energy gap which is free of sub-gap states. Furthermore, the spin splitting of extended Andreev bound states in a short hybrid is observed. 
\\[2ex]
This evidences that the spin mixing from Pt does not adversely affect the semiconducting properties of a hybrid. Crucially, like Al, the Al/Pt system satisfies all the necessary requirements for investigating Majorana zero modes and topological qubits. What should also be considered is that the fabrication of aluminum/platinum samples can be straightforwardly executed, with minimal modifications of the well-established aluminum technology. Importantly, aluminum and platinum are non-toxic materials suited for most UHV deposition chambers. As a result, the development of scalable quantum systems can be readily implemented using Al/Pt bilayers - which is still a major challenge for heavy elements with a low melting point like Sn and Pb. Thus, we expect that Al covered with Pt will be the natural successor to Al-based hybrids. Furthermore, since Al can be grown especially thin in planar geometries, we anticipate that Al/Pt will be particularly attractive for proximitizing two-dimensional semiconductors and van der Waals materials. Future works involving Al/Pt hybrids will focus on investigating Majorana physics, exploring their behavior as a function of chemical potential and high Zeeman energies. 

\section*{Data Availability and Code availability}
Raw data presented in this work and the data processing/plotting codes are available at \\ \url{https://doi.org/10.5281/zenodo.5835794}. Theory simulation code is available upon reasonable request.
\newpage
\section*{Acknowledgements}
We thank Andrei Antipov, Philippe Caroff, William Cole, Dmitry Pikulin, Anton Akhmerov and Roman Lutchyn for helpful discussions. We also thank Jan Cornelis Wolff, Mark Ammerlaan, Olaf Benningshof and Jason Mensingh for valuable technical support. We are grateful to Kongyi Li and Mariusz Andrzejczuk for the help with the TEM lamella preparation. This work has been financially supported by the Dutch Organization for Scientific Research (NWO), the Foundation for Fundamental Research on Matter (FOM) and Microsoft Corporation Station Q.

\section*{Author Contribution}
G.P.M. and N.v.L. conceived the experiment. G.P.M. and N.v.L. fabricated and measured the devices. T.D. and G.W. performed the measurements and analysis of device D. J.Y.W., R.C.D., F.B., P.V., and S.H. assisted with sample fabrication and/or measurements. G.P.M. and N.v.L. analyzed the transport data. A.K. and provided theoretical modelling and analysis of the film and tunnel junction data. S.K. performed the TEM analysis. G.B., S.G. and E.P.A.M.B. carried out the nanowire synthesis. G.P.M. and N.v.L. wrote the manuscript with valuable input from all authors. L.P.K. supervised the project.

\bibliography{scibib}

\bibliographystyle{Science}

\end{document}



\baselineskip24pt


\maketitle


\newpage
\section*{Methods and materials}

\subsection*{Platinum thickness calibration}
\begin{figure}[ht!]
	\includegraphics[width=\textwidth]{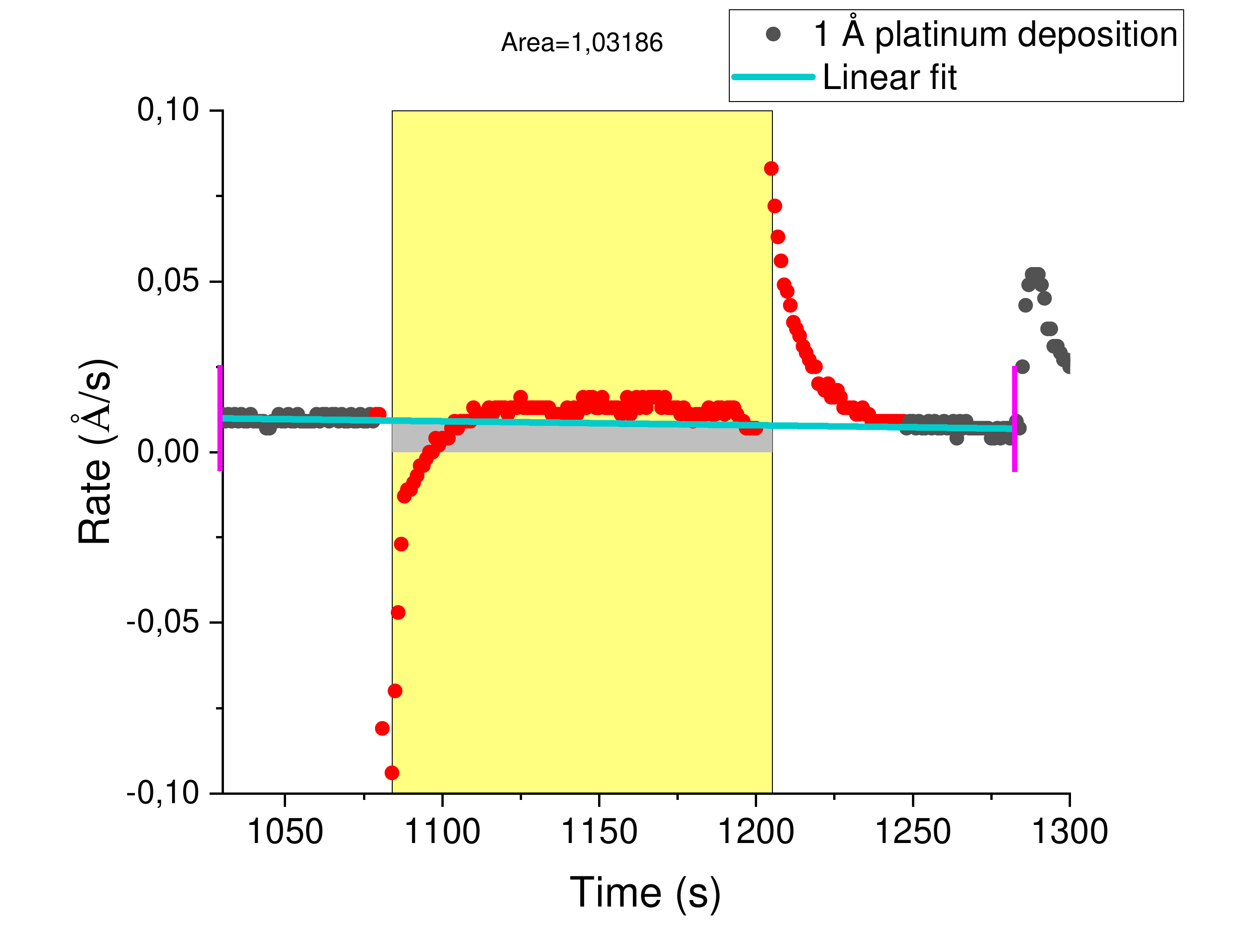}
	\caption{Example of a fitting procedure to precisely estimate the platinum thickness. Dark gray points correspond to the raw rate measured by the quartz crystal monitor. Red points are masked and excluded from the fit. Purple markers indicate the range of the fit, and the cyan line is the fitted linear curve. The yellow rectangle marks the range of integration and the integrated light gray area corresponds to the evaporated platinum thickness.}
\label{fig:Fig_s1}
\end{figure}

In this work we use sub-nanometric amounts of platinum to enhance the magnetic field resilience of aluminum. As platinum and aluminum do not form an abrupt interface, the thickness estimation from TEM analysis is not reliable. Instead, atom-force microscopy (AFM) was used to calibrate the tooling factor of the quartz crystal balance. Prior to the deposition of platinum, the deposition rate was stabilized between a value of 0.01 \AA/s to 0.03 \AA/s. Upon opening of the shutter, there is an apparent sharp drop of the deposition rate due to thermal effects on the quartz crystal monitor (QCM). The rate relaxes back to its initial value after few tens of seconds when the monitor re-cools again. This thermal effect makes it difficult to accurately estimate the thickness of deposited metal during the deposition. To precisely estimate how much Pt was deposited, the deposition rate as a function of time was recorded. Knowing the rate before opening the sample shutter, we calculated the deposition time to obtain the desired thickness. This way we minimized the impact of the thermal effect caused on the estimated thickness by the QCM. After the deposition is complete, the closing of the shutter induces a similar spike on the rate. The rate was allowed to stabilize again at this point. The exact amount of Platinum deposited was then calculated using a linear fit between the stabilized rate before and after the deposition, which accounts for possible drifts in the deposition rate. The curves were then integrated between the opening and closing of the shutter, with the area under the curve corresponding to the deposited Pt thickness. This process is presented in Fig.~\ref{fig:Fig_s1}. 

\subsection*{Al films}
The Al films presented in this work have been deposited with a rate of 0.05\,\AA/s on an intrinsic silicon substrate with 285\,nm thermally oxidized SiO$_{2}$. Prior to the aluminum deposition, the substrate was kept at 138\,K for 1 hour to achieve good thermalization. Metals were evaporated at an angle of $30^{\circ}$ from the normal. After aluminum deposition, platinum was deposited with the above mentioned procedure. AlOx was deposited at 0.1-0.2\,\AA/s. All materials were deposited at a substrate temperature of 138\,K. The critical temperature and magnetic field of the films was measured in a 4-terminal geometry, with 4 probes in a line separated by 1\,$\mu$m. The thickness of the films is 6\,nm as estimated from the quartz crystal monitor (QCM) and transmission electron microscopy (TEM). The films are subsequently capped with a layer of evaporated AlO$_\textrm{x}$, which protects them from oxidation when exposed to air. We note however that the observed thickness of platinum from TEM appears larger than indicated by QCM in the deposition chamber.

The tunnel junctions in this work have been grown under similar conditions. A 6\,nm Al film is deposited at 138\,K with $30^{\circ}$ angle from the normal. The films are oxidized at 138\,K with an oxygen pressure of 10\,Torr. Without breaking the vacuum, 40\,nm Ag is evaporated at 0.5\,\AA/s as the counter electrode at $-30^{\circ}$ from the normal. The samples are subsequently protected with an AlO$_\textrm{x}$ capping layer. They are measured in a 4-terminal geometry, with 2 probes providing the source and drain contacts and the other 2 measuring the voltage drop on the junction.

\begin{figure}[hb!]
	\includegraphics[width=\textwidth]{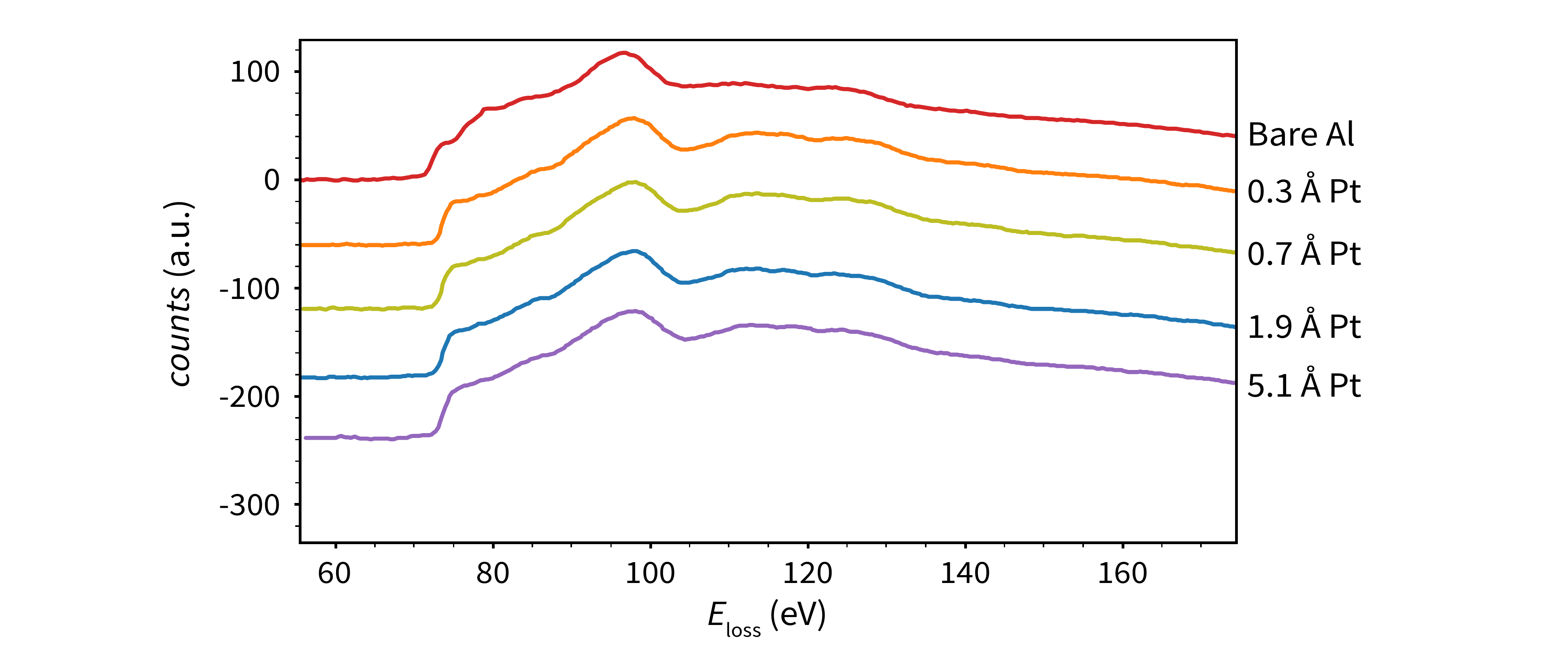}
	\caption{Electron energy-loss spectroscopy extracted at the center of the studied films. The spectra do not change with the increase of Pt thickness. This indicates that Al and Pt do not form a compound, despite the detection of Pt inside the Al layer with EDX. Based on the shape of the spectra we conclude that Pt is physically implanted inside the Al layer during deposition. Values on the y axis have been offset by -60\,a.u. for clarity.}
\label{fig:Fig_s10}
\end{figure}

\newpage
\subsection*{Usadel equation and theory fitting}

We use the Usadel equation \cite{Usadel1970} to calculate the critical fields of the films and fit the tunneling spectroscopy data. The Usadel equation is written in terms of the quasiclassical Green's function \cite{Eilenberger1968,Larkin1969} $\check g(i\omega_n,\boldsymbol r)$ and is valid in the limit of dirty superconductors, $l_{\rm mfp}\ll \xi$, where $l_{\rm mfp}$ is the electronic mean free path and $\xi$ is the superconducting coherence length. Introducing Pauli matrices $\boldsymbol{\hat\sigma}$($\boldsymbol{\hat\tau}$) in spin (Nambu) space, the Usadel equation for a superconductor with spin-orbit impurities in external magnetic field can be written as
\begin{equation}
    D\boldsymbol{\partial}\cdot (\check g\boldsymbol{\partial}\check g)-\left[\omega_n\hat\tau_z+i\boldsymbol{V}_\text{Z}\cdot\boldsymbol{\hat\sigma}\hat\tau_z+\Delta\hat\tau_++\Delta^{\ast}\hat\tau_-+\check\Sigma_{\rm SO},\check g \right]=0,
    \label{eq:Usadel_gen}
\end{equation}
where the covariant derivative is defined as $\boldsymbol{\partial}\boldsymbol{\cdot}=\boldsymbol{\nabla}-i[\boldsymbol{A}\hat\tau_z,\boldsymbol{\cdot}]$, $\boldsymbol{A}$ is the vector potential, $\omega_n$ are Matsubara frequencies, $\boldsymbol{V}_\text{Z}=g_{\rm el}\mu_{\rm B}\boldsymbol{B}/2$ is the Zeeman field originating from the external magnetic field $\boldsymbol{B}$, $g_{\rm el}$ is the electronic $g$-factor, $\mu_{\rm B}$ is the Bohr magneton, $\Delta$ is the superconducting pair potential, and $\hat\tau_{\pm}=(\hat\tau_x\pm\hat\tau_y/2)$. The diffusion constant $D$ corresponds to scattering on non-magnetic impurities and is given by $D=v_{\rm F}l_{\rm mfp}/3$, where $v_{\rm F}$ is Fermi velocity in the superconductor. The quasiclassical Green's function is subject to a normalization condition $\check g^2=1$. As it was first shown in Ref.~\cite{Alexander_PRB:1985}, scattering on spin-orbit impurities produces the self-energy term $\check\Sigma_{\rm SO}=\boldsymbol{\hat\sigma}\check g \boldsymbol{\hat\sigma}/(8\tau_{\rm SO})$ in the Usadel equation, where $\tau_{\rm SO}$ is scattering time. For convenience, we introduce the spin-orbit scattering energy as $\Gamma_{\rm SO}=3\hbar/(2\tau_{\rm SO})$. In general, the Usadel equation has to be supplemented with appropriate boundary conditions, which for a superconductor-insulator interface read $\boldsymbol{\partial}\check g|_{\rm interface}=0$, and the resulting boundary problem has to be solved. However, for very thin superconductors (with thickness $d_{\rm SC}\ll \xi,\lambda_{\rm London}$, where $\lambda_{\rm London}$ is the London penetration depth) in a parallel magnetic field the spatial dependencies of the Green's function and the order parameter can be neglected \cite{deGennes1999}, and the order parameter can be chosen real. In that case, the Usadel equation \eqref{eq:Usadel_gen} becomes an algebraic equation:
\begin{equation}
    \left[\omega_n\hat\tau_z+i\boldsymbol{V}_\text{Z}\cdot\boldsymbol{\hat\sigma}\hat\tau_z+\Delta\hat\tau_1+\check\Sigma_{\rm SO}+\check\Sigma_{\rm ORB},\check g \right]=0,
    \label{eq:Usadel_hom}
\end{equation}
where orbital effects of the magnetic field lead to an additional contribution to the self-energy:
\begin{equation}
    \check\Sigma_{\rm ORB} = \Gamma_{\rm ORB}\frac{\hat\tau_z\check g\hat\tau_z}{4},
\end{equation}
and the orbital depairing energy is given by
\begin{equation}
    \Gamma_{\rm ORB} = \frac{De^2B^2d_{\rm SC}^2}{3\hbar c^2}.
    \label{eq:orb_depairing}
\end{equation}
Equation~\eqref{eq:orb_depairing} is a familiar result for thin superconducting films subjected to a parallel magnetic field \cite{Maki1964,Tinkham2004}.

To calculate the order parameter self-consistently, Eq.~\eqref{eq:Usadel_hom} needs to be solved together with the gap equation,
\begin{equation}
    \Delta \log\left(\frac{T}{T_{c0}}\right)=2\pi T\sum_{\omega_n>0}\left( \frac{1}{4}\text{Tr}(\hat\tau_x\check g)-\frac{\Delta}{\omega_n} \right),
    \label{eq:gap_eq}
\end{equation}
where $T$ is temperature and $T_{c0}$ is critical temperature of the bare superconductor. In addition, to appropriately identify first-order transitions one needs to ensure that the free energy difference between the superconducting and the normal state is negative throughout the calculation \cite{Aikebaier2019,Heikkila2019,Khindanov2021}. Once the self-consistent value of the order parameter and the corresponding Green's function are obtained, the density of states in the superconductor can be calculated using 
\begin{equation}
    N(E)=\frac{1}{8}N_0\text{Re}[\text{Tr}(\hat\tau_z\check g|_{\omega_n\to-iE^+})],
    \label{eq:dos}
\end{equation}
where $N_0$ is the density of states at the Fermi level. The differential conductance in the SIN junction is related to the density of states through convolution \cite{Meservey_1994:PR},
\begin{equation}
    \frac{dI}{dV} (V)\propto \int_{-\infty}^{\infty} N(E)K(E+eV)dE,
    \label{eq:cond}
\end{equation}
where $V$ is a voltage bias and the convolution kernel is given by
\begin{equation}
    K(x)=\frac{\beta e^{\beta x}}{(1+e^{\beta x})^2}
\end{equation}
with the inverse temperature $\beta=1/k_{\rm B}T$.

\subsection*{Simulation details}
We use Eqs.~\eqref{eq:Usadel_hom}-\eqref{eq:gap_eq} to calculate the values of the pair potential and to simulate the critical fields of the Al and Al/Pt films considered in Fig.~1 of the main text. We take  the critical temperature to be $T_{c0}=1.79$ K for all simulated samples. Using the critical field of the bare Al film (which we assume has a negligible amount of spin-orbit impurities) measured in the experiment, $B_c\approx 2.6$ T, we extract the value of the mean free path in the film by solving the Usadel equation \eqref{eq:Usadel_hom}-\eqref{eq:gap_eq} with $\check\Sigma_{\rm SO}=0$ and obtain $l_{\rm mfp}\approx 0.9$ nm. We further use this value to simulate the critical fields of the Pt-covered samples (see red dashed curve in Fig.~1C of the main text) and extract the respective values of the spin-orbit scattering rate (see Fig.~\ref{fig:Fig_sX}A). Values of the parameters used in these simulations, both independent and extracted by fitting experimental data, are given in Table~\ref{tab:params1}.
\begin{table}[]
    \centering
    \begin{tabular}{|c|c|c|c|c|c|c|c|c|c|c|}
    \hline
    \multicolumn{4}{|c|}{Independent parameters} & \multicolumn{3}{|c|}{Extracted Parameters} \\
\hline
$T$\,(mK) & $d_{\rm SC}$\,(nm) & $g_{\rm el}$ & $v_{\rm F}$\,(m/s) & $\Delta_{0}$\,(meV) & $l_{\rm mfp}$\,(nm) & $\Gamma_{\rm SO}$\,(meV)  \\
\hline
20 & 6 & 2 & $2\times 10^6$ & 0.27 & 0.9 & see Fig.~\ref{fig:Fig_sX}A \\
\hline
    \end{tabular}
    \caption{Values of parameters used in critical field simulations of Al and Al/Pt films (see Fig.~1 of the main text). Extracted parameters were obtained by fitting experimental data.}
    \label{tab:params1}
\end{table}

Conductance spectroscopy on the Al and Al/Pt films presented in Fig.~2 of the main text is simulated using Eqs.~\eqref{eq:Usadel_hom},\eqref{eq:gap_eq},\eqref{eq:dos} and \eqref{eq:cond}. Table~\ref{tab:params2} summarizes values of the parameters, both independent and extracted by fitting experimental data, used in conductance simulations.

\begin{table}[]
    \centering
    \begin{tabular}{|c|c|c|c|c|c|c|c|c|c|c|c|}
    \hline
    \multicolumn{1}{|c|}{} & \multicolumn{4}{|c|}{Independent parameters} & \multicolumn{3}{|c|}{Extracted Parameters} \\
\hline
& $T$\,(mK) & $d_{\rm SC}$\,(nm) & $g_{\rm el}$ & $v_{\rm F}$\,(m/s) & $\Delta_{0}$\,(meV) & $l_{\rm mfp}$\,(nm) & $\Gamma_{\rm SO}$\,(meV) \\
\hline
Al  & 110 & 4.5 & 2 & $2\times 10^6$ & 0.31 & 0.68 & 0.0 \\
\hline
Al/Pt  & 30 & 4.5 & 2 & $2\times 10^6$ & 0.306 & 0.68 & 7.5 \\
\hline
    \end{tabular}
    \caption{Values of parameters used in conductance spectroscopy simulations of Al and Al/Pt tunnel junctions (see Fig.~2 of the main text). Extracted parameters were obtained by fitting experimental data.}
    \label{tab:params2}
\end{table}

Similarly to initial studies of the Al/Pt system, our theoretical model yields a linear dependence of $\Gamma_{\rm SO}$ on $d_{\rm Pt}$. We note however that even for $d_{\rm Pt}$ = 5.1\,\AA, the extracted value of $\Gamma_{\rm SO}$ is smaller than $\Gamma_{\rm SO}$ = 7.5\,meV, which we extract from the tunneling measurements. Initial studies of Al/Pt revealed unphysically large spin-orbit scattering rates~\cite{Tedrow_1982:PRB}. In the case of the above mentioned experiment, the extracted spin-orbit scattering rate was higher than the momentum scattering rate, which indicated that the increase of critical field was not fully understood. It was pointed out later~\cite{Alexander_PRB:1985} that, due to Fermi liquid effects, the $g$-factor of such a thin Al films is being reduced~\cite{Meservey_1994:PR}. We plot the energy difference between the spin-up and spin-down quasiparticle peaks in a 4.5-nm Al film as a function of magnetic field in Fig.~\ref{fig:Fig_sX}B. The analysis is made for fields larger then 1\,T. In our case, we do not observe a clear deviation from a $g$-factor of 2 (indicated by the orange curve). The slight discrepancy observed near the transition can be a result of the peak broadening, rather than Fermi-liquid effects. We note, however, that the Fermi-liquid correction becomes more relevant at higher magnetic field values, and we cannot fully exclude the presence of these effects in Al/Pt devices. 

\vfill
\begin{figure}[h!]
	\includegraphics[width=\textwidth]{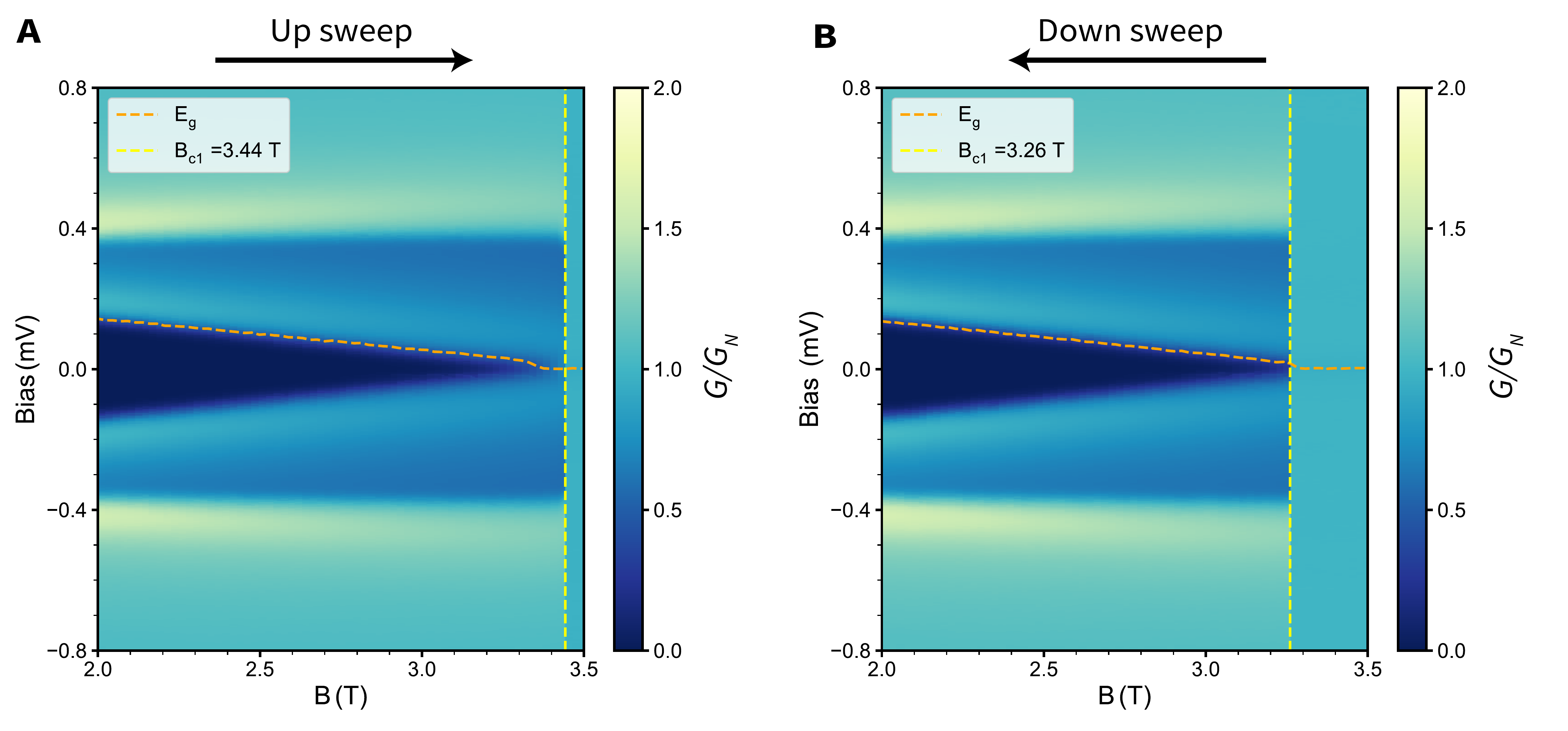}
	\caption{Additional experimental tunnel junction data in the vicinity of the critical field of the Al tunnel junction. Panels illustrate the magnetic field up (\textbf{A}) and down (\textbf{B}) sweeps, respectively. Measurements show a clear difference in the critical field value, evidencing hysteresis as expected for the first order phase transition. Orange lines depict the gap edge and yellow lines mark the critical field extracted from experimental data.}
\label{fig:Fig_Sx_hysteresis}
\end{figure}

\begin{figure}[h!]
	\includegraphics[width=\textwidth]{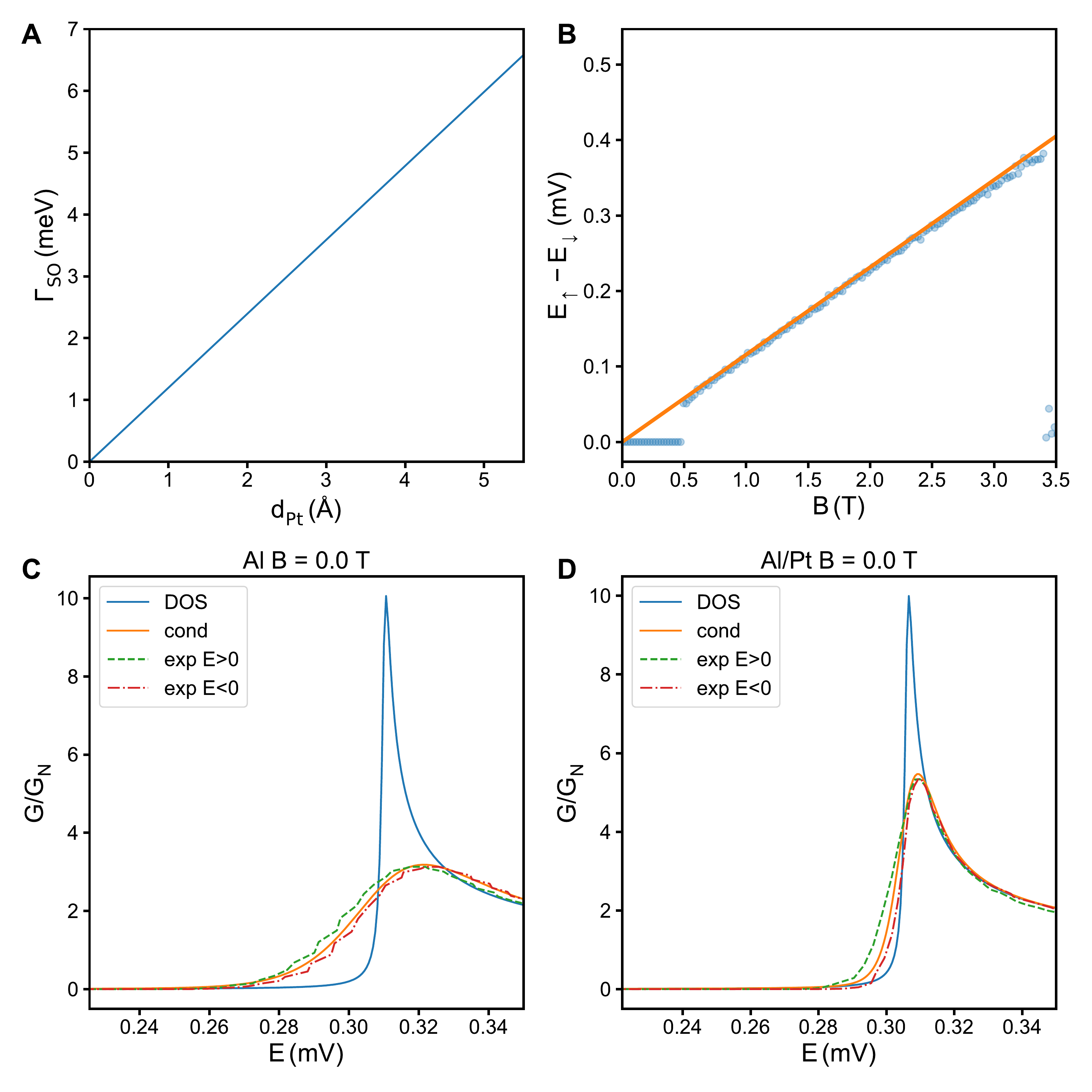}
	\caption{Panel (\textbf{A}) depicts the value of the spin-orbit scattering rate $\Gamma_{\rm SO}$ as a function of platinum thickness fitted to experimental film data presented in the main text. (\textbf{B}) Experimentally extracted energy difference between Zeeman-split quasiparticle peaks (blue dots), and the corresponding Zeeman energy with $g=2$. (\textbf{C}) Calculated DOS (blue) and thermally broadened conductance (orange) for the Al tunnel junction. Green and red curves correspond to positive and negative bias respectively. (\textbf{D}) Same curves for the Al/Pt tunnel junction.}
\label{fig:Fig_sX}
\end{figure}

\newpage
\subsection*{Nanowire hybrids: substrate fabrication}
The nanowire hybrids presented in this work are fabricated on pre-patterned substrates, following the shadow-wall lithography technique described in \cite{Heedt:2021_NC,Borsoi:2021_AFM}. Intrinsic silicon wafers ($2\, k\Omega\,\cdot\,$cm) with $285\,$nm thermal SiO$_\textrm{x}$ serve as the base for the device substrates. Local bottom gates are patterned with standard electron-beam litography (EBL) techniques, using PMMA 950k A2 spun at $4\,$krpm for one minute followed by 10 minutes of baking on a $185\, ^{\circ}$C hot plate. After development of the resist using a 3:1 solution of IPA and MIBK, $3\,$nm Ti and $17\,$nm Pd are deposited as the gate metal using e-beam evaporation at $0.5 \,\AA$/s and $1\, \AA$/s, respectively. Subsequently, bond pads are patterned with EBL using PMMA 950k A6 spun at $4\,$krpm for one minute and hot-baked at $185\, ^{\circ}$C for 10 minutes. After development, $50\,$nm of W is sputtered  using RF-sputtering at $150\,$W in an Ar pressure of $20\,\mu$bar. Next, the substrates are covered with high-quality HfO$_x$ gate dielectric grown at $110\, ^{\circ}$C using atomic layer deposition (ALD). Finally, shadow walls are patterned on top of the dielectric. FOx-25 (HSQ) is spun at $1.5\,$krpm for one minute, followed by 2 minutes of hot baking at $180\, ^{\circ}$C and patterning with EBL. The HSQ is then developed with MF-321 at $60\, ^{\circ}$C for 5 minutes and the substrates are subsequently dried using critical point dryer (CPD). Nanowires are deposited onto the gates using an optical nanomanipulator setup. For some devices, a modified version of the fabrication flow presented above was used.
\\
For devices H and I, the shadow walls are grown using a double-layer process. First, PMMA 950k A8 is spun at $4\,$krpm and hot baked at $185\, ^{\circ}$C for 10 minutes. After EBL and development, FOx-25 (HSQ) is spun at $2\,$krpm for one minute, followed by 2 minutes of hot baking at $180\, ^{\circ}$C and patterning with EBL. HSQ is developed for 15 minutes using  MF-321 at $60\, ^{\circ}$C, followed by stripping of the PMMA using acetone at  $50\, ^{\circ}$C for 10 minutes and drying of the substrates using CPD.
\\
For device H, bottom gates are instead fabricated using sputtered W, etched down using reactive ion etching (RIE). Details are described in \cite{Heedt:2021_NC,Borsoi:2021_AFM}.
\\
For devices B, H and I, AlO$_x$ grown at $300\, ^{\circ}$C using ALD was used as the gate dielectric instead of HfO$_x$. Since the HSQ developer etches AlO$_x$, the ALD is moved in the fabrication flow to be done after the shadow walls are fabricated.
\\
For device J, p++ doped Si wafers ($0.02\,\Omega\,\cdot\,$cm) with $285\,$nm thermal SiO$_\textrm{x}$ were used. The Si serves as a global back gate, and so no local gates were patterned. Instead, only the bond pads and shadow walls were fabricated onto the substrate.

\subsection*{Nanowire hybrids: superconductor deposition}
To obtain a pristine, oxide-free semiconductor surface, a gentle oxygen removal is accomplished via atomic hydrogen radical cleaning. For this purpose, a custom-made H radical generator is installed in the load lock of an aluminium electron-gun evaporator. It consists of a gas inlet for H$_2$ molecules connected to a mass-flow controller and a tungsten filament at a temperature of about $1700\,^{\circ}$C that dissociates a fraction of the molecules into hydrogen radicals. The optimal removal of the native oxide is achieved for a process duration of $60\,$mins and at a H$_2$ pressure of 6.3$\cdot10^{-5}\,$mbar. This recipe, which is used for all the devices shown in this paper, results in a constant EDX count of oxygen at the interface as shown in the previous works utilizing our shadow wall lithography technique (see Refs\cite{Heedt:2021_NC,Borsoi:2021_AFM}). 

\begin{figure}[h!]
	\includegraphics[width=\textwidth]{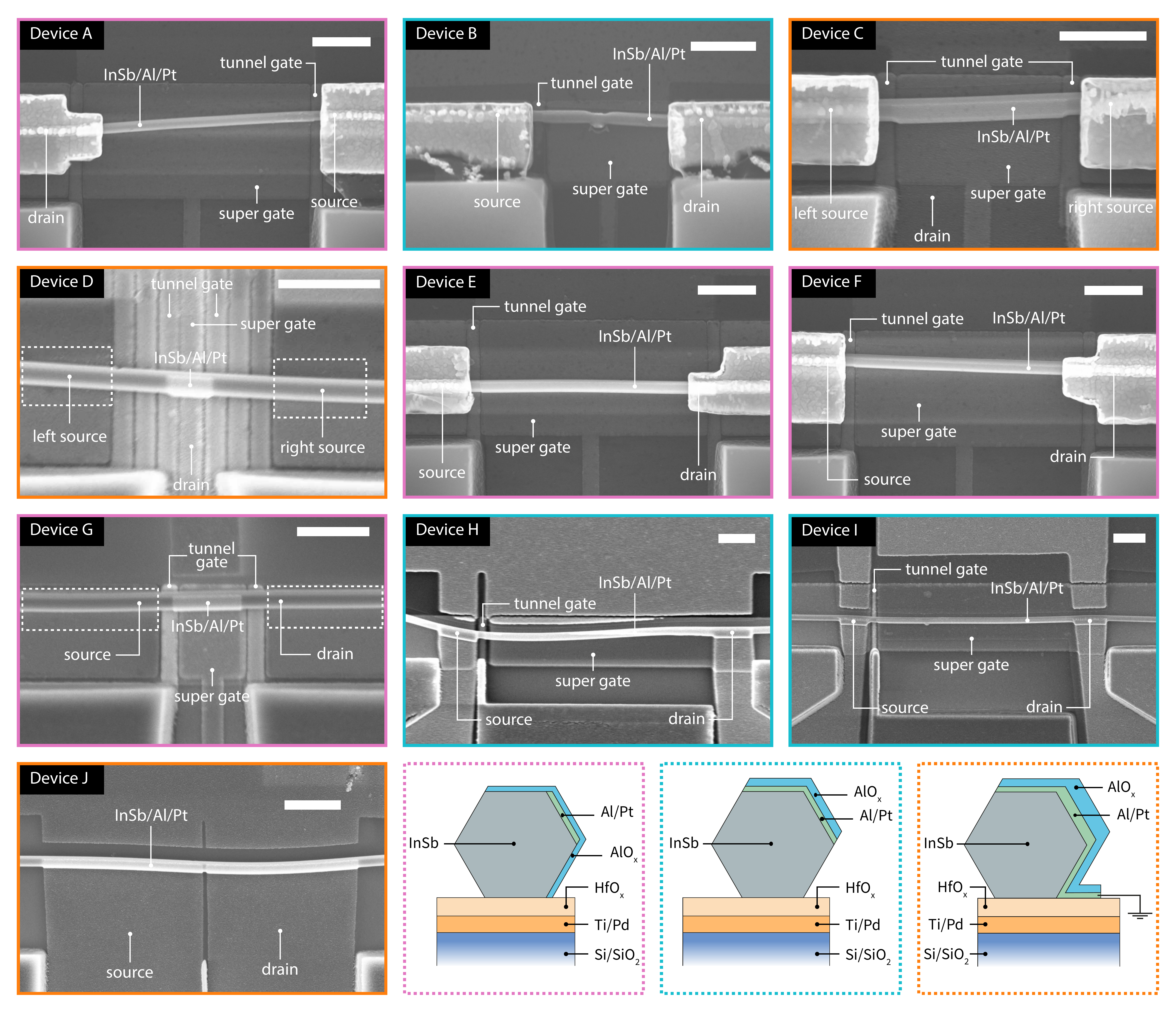}
	\caption{SEM images of the nanowire hybrid devices presented in this work. Scale bars are 500\,nm. The borders of each panel are color-coded corresponding to the three bottom-right panels, which show illustrations of the intended cross-section of various fabrication recipes used in this work. Device B, which showed Coulomb oscillations, was not intended to be an island device; we suspect a second barrier was present at the grounding contact due to incomplete Ar milling of the AlO$_{\rm x}$ capping layer. Additionally, it shows signs of ESD which we suspect happened during unloading from the fridge. The images of devices D and G were taken prior to contact deposition - dashed rectangles indicate the designed contact location.}
\label{fig:Fig_s2}
\end{figure}

After the native oxide removal, the samples are cooled down to 138\,K and thermalized for one hour. The Al is then deposited with a rate of 0.05\,\AA/s. The various samples presented in this work are shown in Fig.~\ref{fig:Fig_s2}, and can be separated into three categories based on the facet coverage of the nanowire. For nanowires with a single covered facet (pink outline in Fig.~\ref{fig:Fig_s2}), a thin Al film is grown with a $30^\circ$ angle with respect to the substrate. After Pt deposition at the same angle, the film is oxidized while still cold at an oxygen pressure of 200\,mTorr for 5 minutes. With these growth conditions, the top and botom-side facets of the nanowire are covered with extremely thin granular Al and are expected to fully oxidize. Similarly, 2 nanowire facets can be covered with Al (cyan outline in Fig.~\ref{fig:Fig_s2}) by growing a slightly thicker film at $50^\circ - 60^\circ$ from the substrate. On the other hand, nanowires with 3-facet coverage (orange outline in Fig.~\ref{fig:Fig_s2}) can be grown at $30^\circ$ (or a mix of $15^\circ$ and $45^\circ$) with thicker films. In that case, there is a continuous connection between the nanowire shell and the film on the substrate. This can serve as a ground or source/drain contact to the nanowire. These samples are typically capped with evaporated AlO$_\textrm{x}$ ($\sim$\,0.2\,\AA/s) to prevent oxidation of the shell-substrate connection. Details on the growth conditions of all presented samples are shown in Table~\ref{tab:table1}.

\subsection*{Nanowire hybrids: contacts}
For most devices presented in this work, contacts are fabricated ex-situ after the superconductor deposition. PMMA 950k A6 is spun at 4\,krpm and subsequently cured at room temperature in a vacuum oven to prevent intermixing at the pristine InSb-Al interface. Contacts are patterned using EBL, and Ar milling is used to remove the native oxide prior to deposition of 10\,nm Cr and 120\,nm Au using e-beam evaporation at 0.5\,$\textrm{\AA}$/s and 1.5\,$\textrm{\AA}$/s, respectively.
\\
For devices H and I, contacts are deposited in-situ. Using the single-shot fabrication technique presented in \cite{Borsoi:2021_AFM}, 50\,nm Pt is deposited at a 30$^\circ$ angle with respect to the substrate to form metallic contacts. 
\\
For device J, the Al/Pt shell is deposited at a 30$^\circ$ angle with respect to the substrate, forming the source and drain contacts.  

\subsection*{Measurements}
Transport measurements are conducted in dilution refrigerators with a base temperature of $\sim$ 20\,mK. All magnetic field measurements presented in this work have the magnetic field aligned parallel to the nanowire using 3-axis vector magnets. Voltage-bias measurements were conducted in a 2-terminal geometry (with the exception of devices C and D) using standard lock-in techniques. The used excitation voltages are between 10$\,\mu$V and 20$\,\mu$V, with frequencies between 15$\,$Hz and 40$\,$Hz. To calculate the voltage drop on the sample and the corresponding conductance, a setup-specific series resistance is taken into account (see Table \ref{tab:table1}). The used series resistances consist of the input resistance of the current amplifier, output resistance of the voltage source and resistance of the RC filters present in the dilution fridge (no contact resistance is assumed in any of the measurements).
Measurements in the 3-terminal geometry on devices C and D are conducted using the circuit described in \cite{Martinez:2021_Arxiv}. For the data presented on device C, only the DC voltage drop on each junction is corrected. Measurements taken on device D are not corrected for any series resistances.
To extract the induced gap size from measurements on device C, each voltage-bias line trace is first smoothed by applying a Savitzky-Golay filter. Subsequently, the data is split into positive bias and negative bias, before being normalized by the peak non-local conductance value in the corresponding half. The induced gap is then determined separately for positive and negative bias values, by calculating the voltage value for which the signal crosses 20\% of the peak value. The energy gap $E_{\rm g}$ shown in the main text Fig.~4B is the average of these two values. See Fig~\ref{fig:Fig_s9} for the non-local data with an overlay of the extracted $E_{\textrm{g}}$ values, as well as the processed data with overlay. The non-local slope is calculated by averaging the gradient of the Savitzky-Golay filtered data within a $\pm\,9\,\mu$V window around zero bias.

\vfill
\begin{table}[h!]
    \centering
    \begin{tabular}{c|c|c|c|c|c||c|c}
        \hline
         \textbf{Device} & \textbf{\# facets} & \textbf{Angle ($^{\circ}$)} & \textbf{d$_{\rm Al}$ (nm)} & \textbf{Oxidation} & \textbf{d$_{\rm Pt}$ ($\AA$)} & \textbf{Figure} &  \textbf{R$_{\rm series}$ ($\Omega$)}  \\
         \hline
         A & 1 & 30 & 5.5 & $200\,$mTorr O$_2$ & 1.8 & 3D & 15134\\ 
           &   &   &   &   &    & 3E & 15134\\
           &   &   &   &   &    & 3F & 15134\\
           &   &   &   &   &    & S6 & 15134\\
           &   &   &   &   &    & S7 & 6134\\
         B & 2 & 60 & 7.5 & AlO$_x$ capping & 2 & 3B & 6134\\
           &   &   &   &   &    &   3C & 6134\\
           &   &   &   &   &    &   S10 & 6134\\
         C & 3 & 15 + 45 & 4 + 4 & AlO$_x$ capping & 2 & 4 & 6778\\
         D & 3 & 15 + 45 & 4 + 4 & AlO$_x$ capping & 2 & 5 & 0\\
         E & 1 & 30 & 5.5 & $200\,$mTorr O$_2$ & 1.8 & S8 & 15134\\
         F & 1 & 30 & 5.5 & $200\,$mTorr O$_2$ & 1.8 & S9 & 6134\\
         G & 1 & 30 & 8 & $200\,$mTorr O$_2$ & 1.8 & S11 & 6144\\
         H & 2 & 50 & 7.5 & $200\,$mTorr O$_2$ & 2 & S12 & 8668\\
         I & 2 & 50 & 7.5 & $200\,$mTorr O$_2$ & 2 & S13 & 15134\\
         J & 3 & 30 & 12.5 & $200\,$mTorr O$_2$ & 5 & S14 & 6778\\

    \end{tabular}
    \caption{Overview of sample fabrication parameters. The deposition angle is specified with respect to the substrate. The \# facets column indicates the amount of nanowire facets covered with Al/Pt, of which cross-section illustrations are shown in Fig \ref{fig:Fig_s2}. Right two columns show the subtracted series resistance from the raw data for each plot, which consists of the input resistance of the current amplifier, output resistance of the voltage source and resistance of the RC filters present in the dilution fridge. For devices C \& D, we refer the main body of manuscript, as the series resistance varies for the grounding and biasing lines.}
    \label{tab:table1}
\end{table}
\clearpage

\newpage
\section*{Additional data on nanowire hybrids}

\begin{figure}[h!]
	\includegraphics[width=\textwidth]{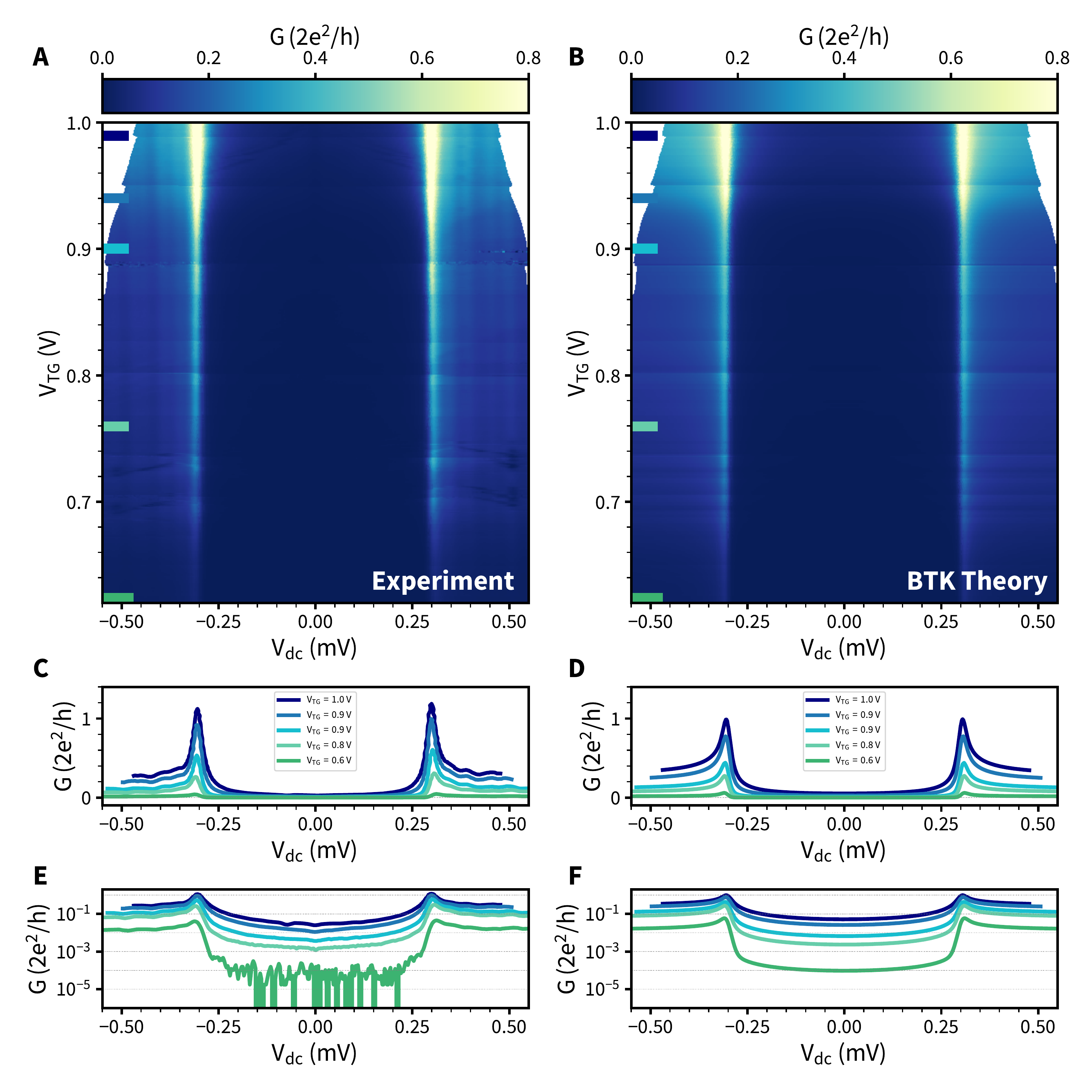}
	\caption{\textbf{Device A} additional data. \textbf{(A)} Differential conductance as a function of tunnel gate voltage, taken at $V_{\rm SG} = -0.8$\,V. \textbf{(B)} BTK theory, with $\Delta$ = 304\,$\mu$eV, $T$ = 50\,mK and transmission $G_{\rm N}$ estimated from the experimental data. Bottom panels show conductance traces at various transmissions of the experiment \textbf{(C)} and theory \textbf{(D)} in a linear scale. The same traces are shown in logarithmic scale in panels \textbf{(E,F)}.}
\label{fig:Fig_s3}
\end{figure}

\begin{figure}[h!]
	\includegraphics[width=\textwidth]{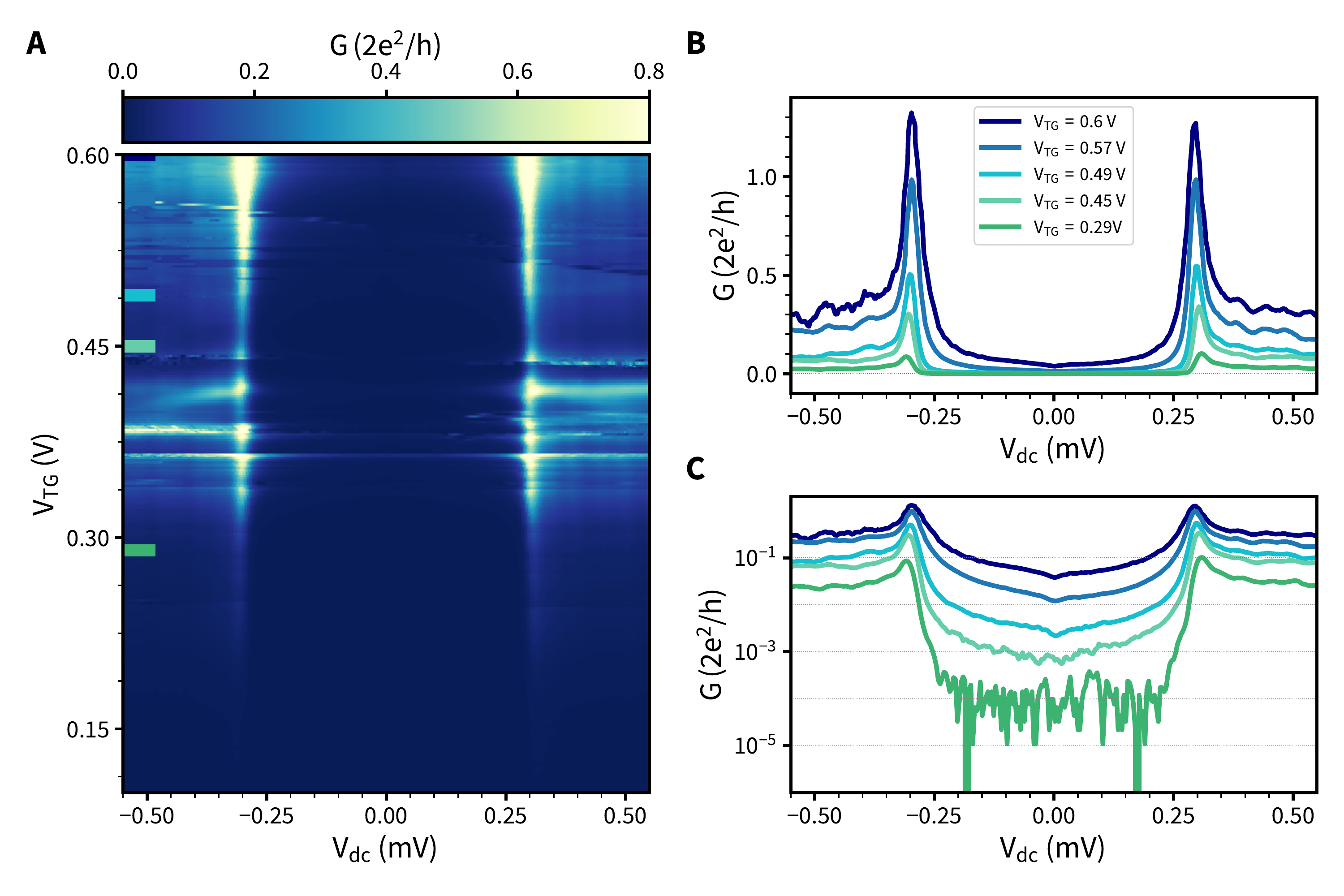}
	\caption{\textbf{Device A} additional data. \textbf{(A)} Differential conductance as a function of tunnel gate voltage, taken at $V_{\rm SG} = -2.0$\,V. At this super gate voltage, an unfavorable electrostatic potential leads to more disordered transport in the tunnel junction. Line cuts taken at different transmission values of the tunnel junction are shown in the linear \textbf{(B)} and logarithmic \textbf{(C)} scale.}
\label{fig:Fig_s3}
\end{figure}

\begin{figure}[ht]
	\includegraphics[width=\textwidth]{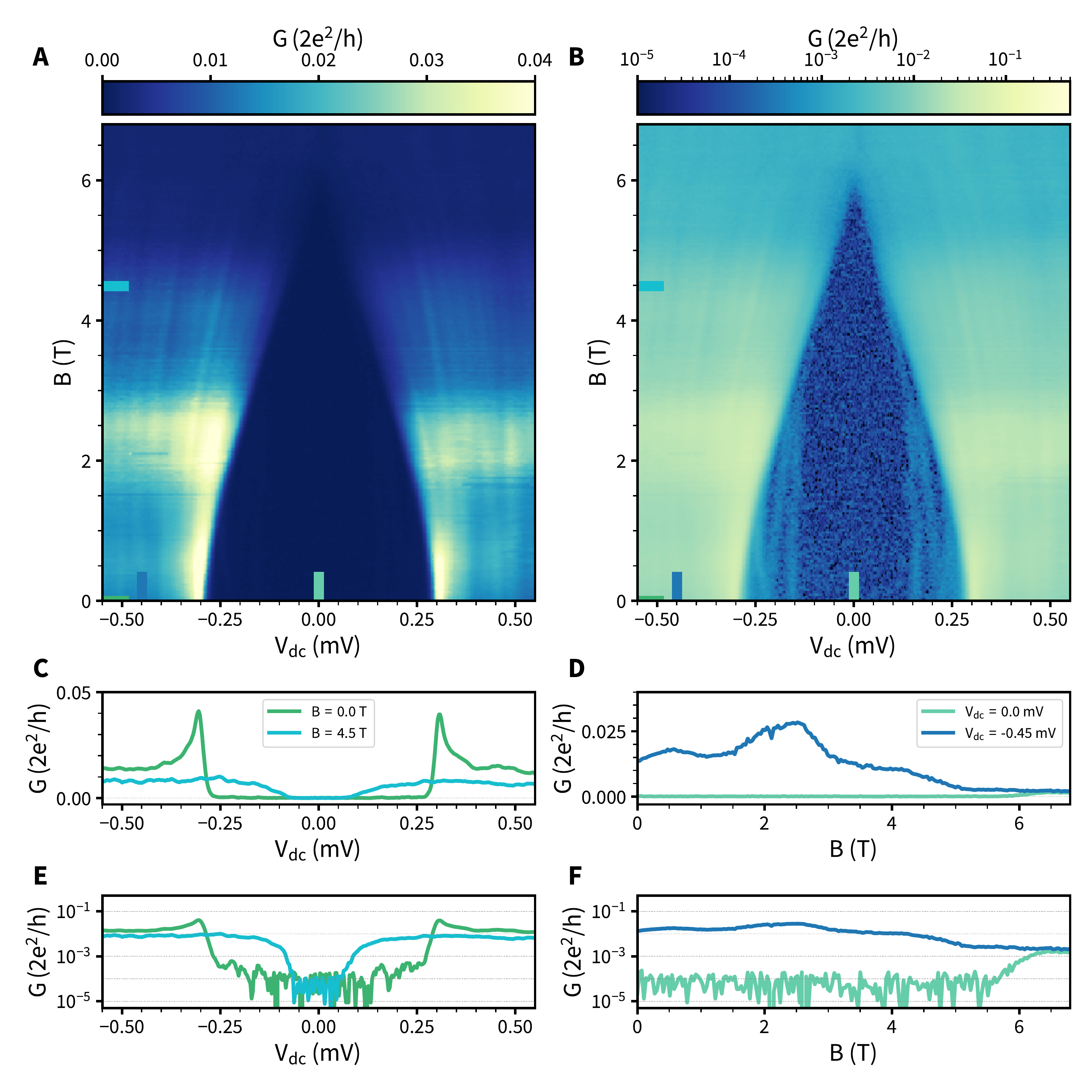}
	\caption{\textbf{Device E.} Additional data on magnetic field compatibility of single-facet devices. Differential conductance vs parallel magnetic field map presented in the linear \textbf{(A)} and logarithmic \textbf{(B)} scale. Panel \textbf{(B)} reveals a few discrete sub-gap states, which typically originate from disorder in the semiconducting junction. Panels \textbf{(C)} and \textbf{(D)} depict linecuts taken at $B$ = 0\,T (green) and $B$ = 4.5\,T (cyan). Out-of-gap and in-gap conductances as a function of magnetic field are shown in the linear \textbf{(D)} and logarithmic scale. The length of the studied device is about 1.8$\,\mu$m, similar to a device presented in the main text in Fig 2. The data is taken at $V_{\rm SG}$ = -2\,V.}
\label{fig:Fig_s4}
\end{figure}

\begin{figure}[ht]
	\includegraphics[width=\textwidth]{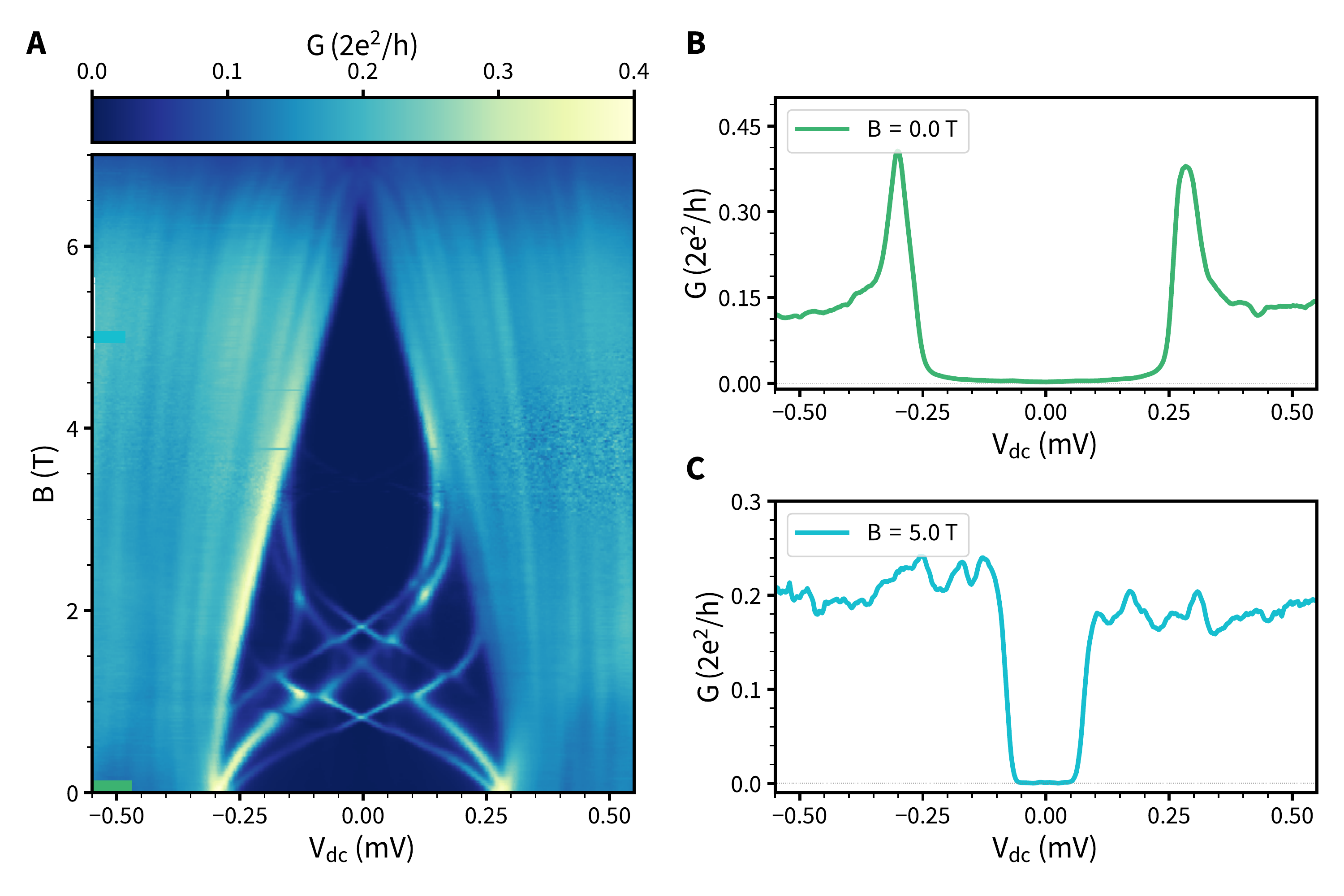}
	\caption{\textbf{Device F.} Additional data on magnetic field compatibility of single-facet devices. Differential conductance as a function of parallel magnetic field \textbf{(A)} at V$_{\rm SG}$ = 0\,V.  Discrete Andreev bound states cross and anti-cross inside the gap. Above $B$ = 4\,T, the gap is hard again. Panels \textbf{(B)} and \textbf{(C)} depict differential conductance scans at $B$ = 0\,T and $B$ = 5\,T.}
\label{fig:Fig_s5}
\end{figure}

\begin{figure}[ht]
	\includegraphics[width=\textwidth]{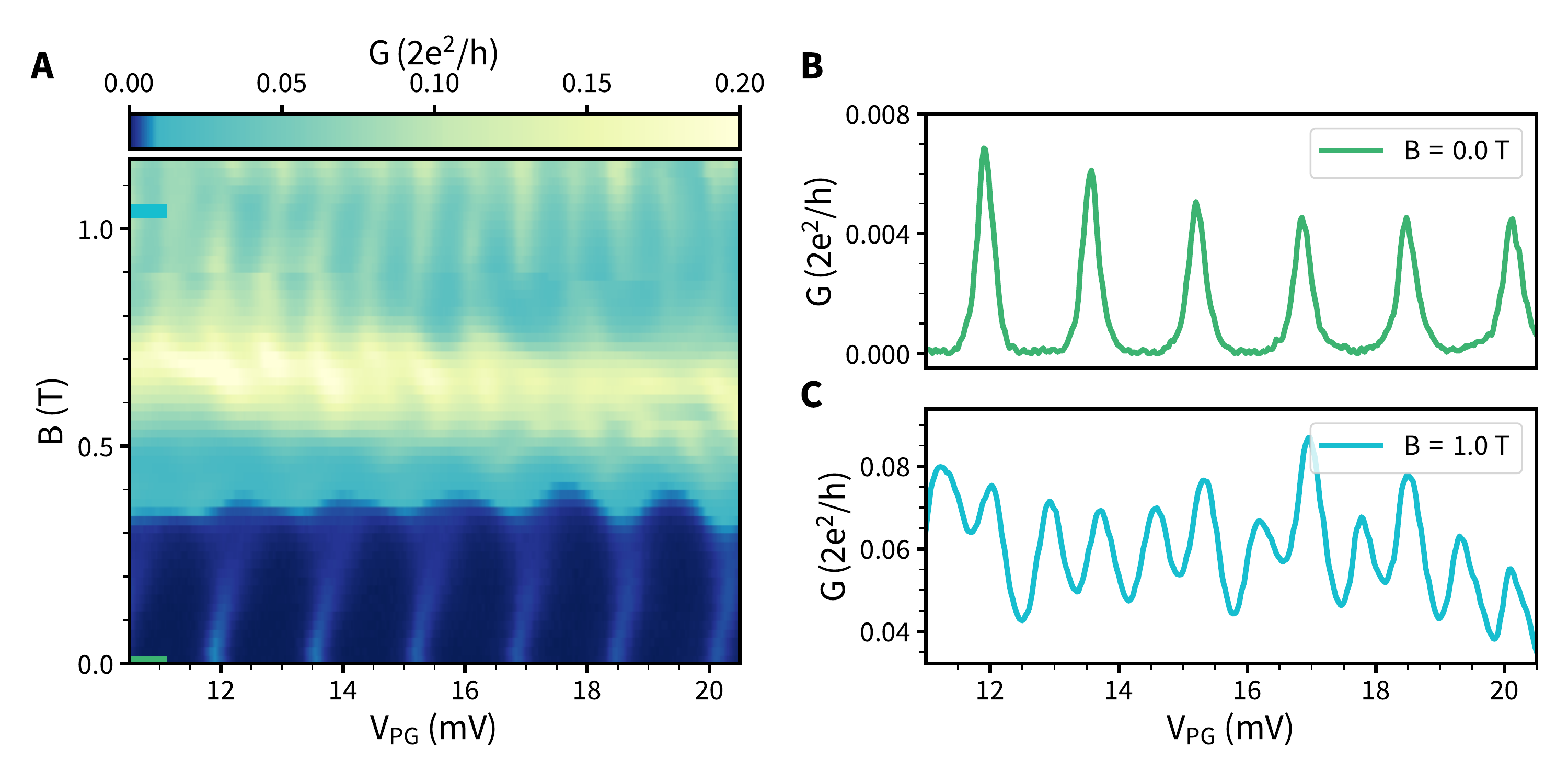}
	\caption{\textbf{Device B.} Magnetic field dependence of 2$e$ transport presented in Fig.~3 in the main text. A 2$e$ to 1$e$ transition is observed when the energy of the lowest energy sub-gap state drops below the charging energy of the island.}
\label{fig:Fig_sx1}
\end{figure}

\begin{figure}[ht]
	\includegraphics[width=\textwidth]{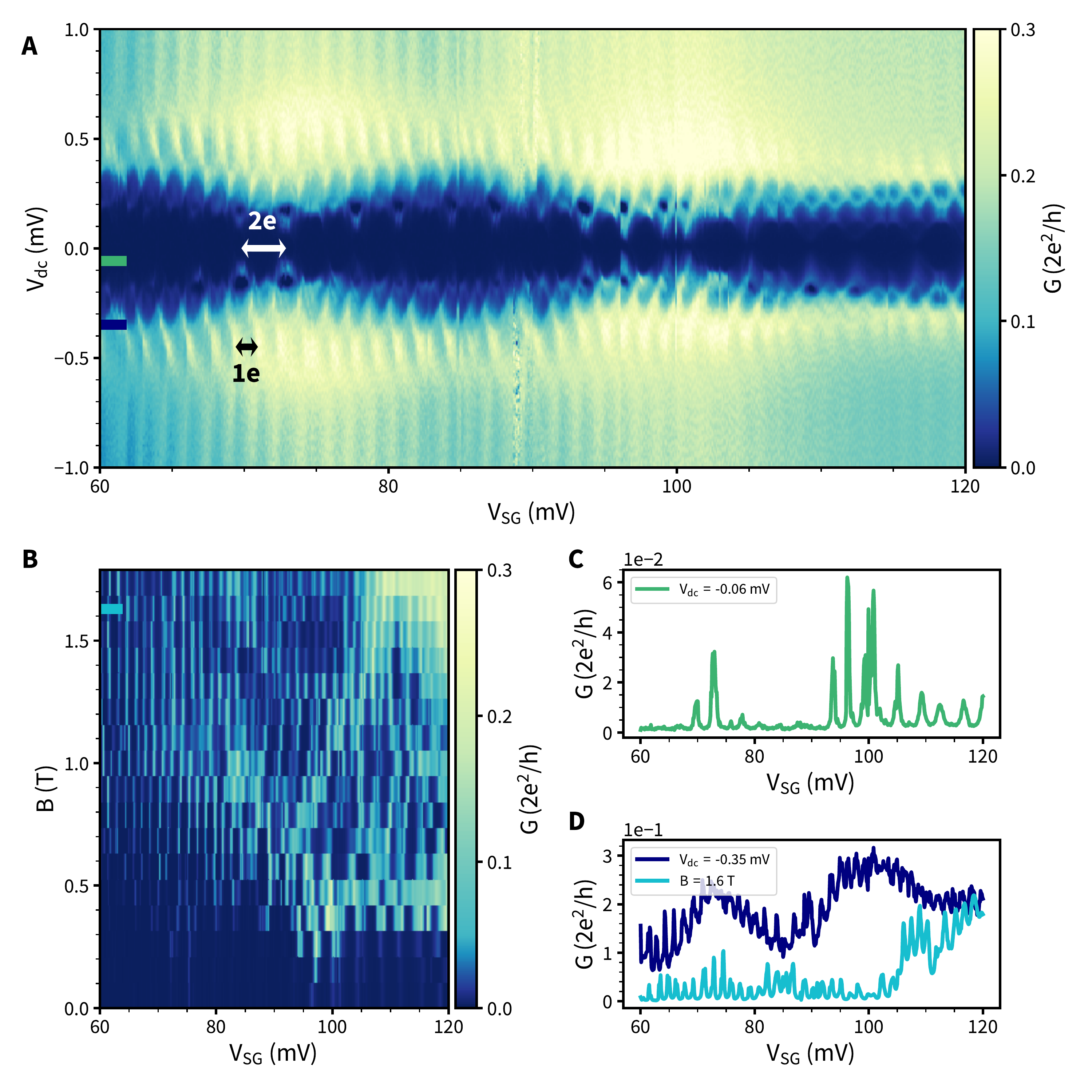}
	\caption{\textbf{Device G.} Additional data presenting 2$e$ charging on a single-facet device. Panel (\textbf{A}) illustrates voltage-bias spectroscopy showing clear 2$e$ and 1$e$ charging processes. Panel (\textbf{B}) illustrates magnetic field dependence of 2$e$ transport. The linecut taken at zero bias is depicted on panel (\textbf{C}). Line cuts at higher bias ($B =$ 0\,T) and $B=$ 1.6\,T (zero bias) are shown in panel (\textbf{D}).}
\label{fig:Fig_sx2}
\end{figure}
\clearpage

\begin{figure}[ht]
	\includegraphics[width=\textwidth]{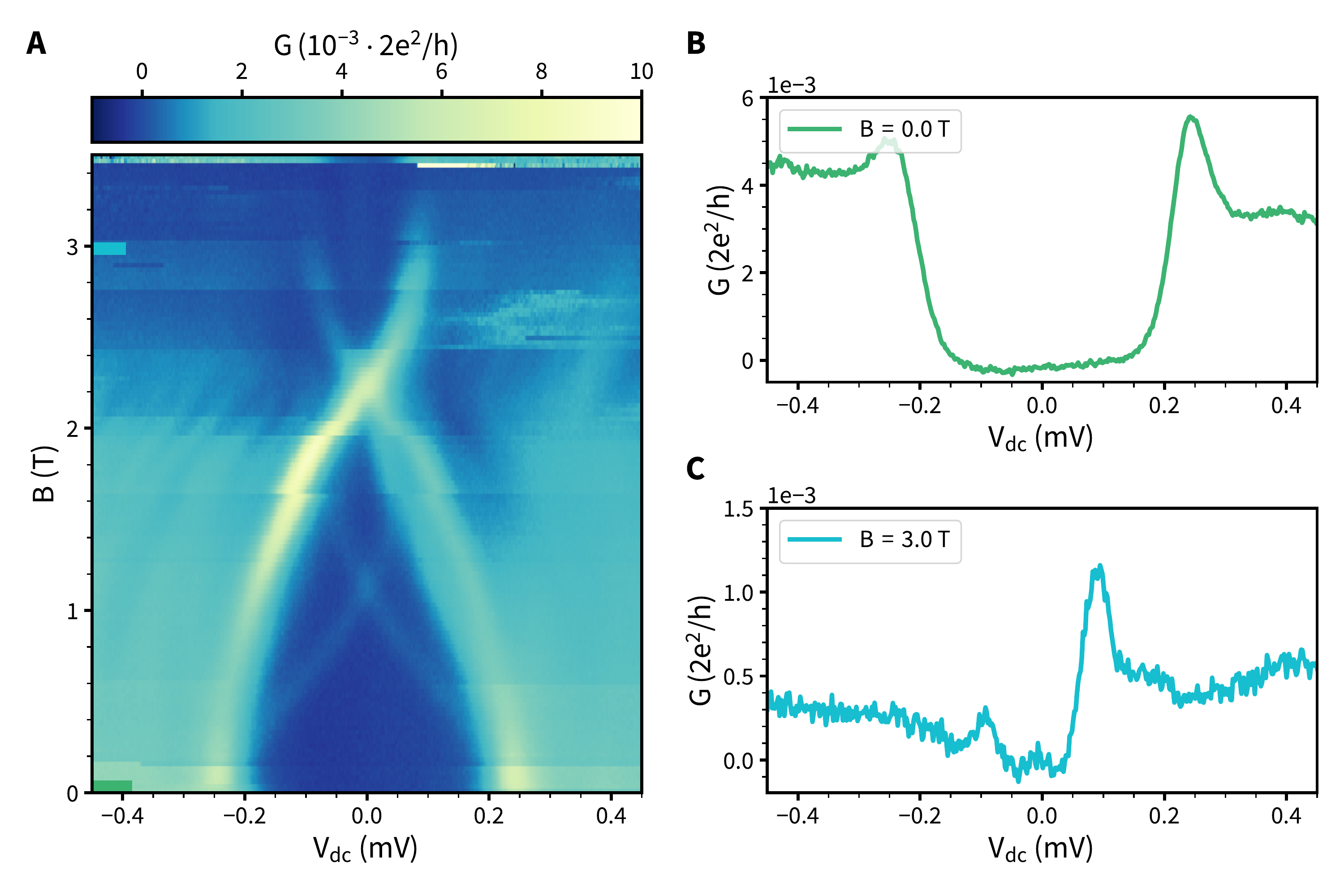}
	\caption{\textbf{Device H.} Example of a  2-facet device fabricated with the single-shot technique. Panel \textbf{(A)} presents the differential conductance map as a function of magnetic field, which resembles the closing and reopening of the superconducting gap followed by the formation of a faint but stable zero-bias peak. Panels \textbf{(B)} and \textbf{(C)} depict differential conductance scans at $B$ = 0\, and $B$ = 3\,T. Due to non-functioning super gate and dielectric instability, the device could not be further investigated.}
\label{fig:Fig_s6}
\end{figure}
\clearpage

\begin{figure}[h]
    \centering
	\includegraphics[width=\textwidth]{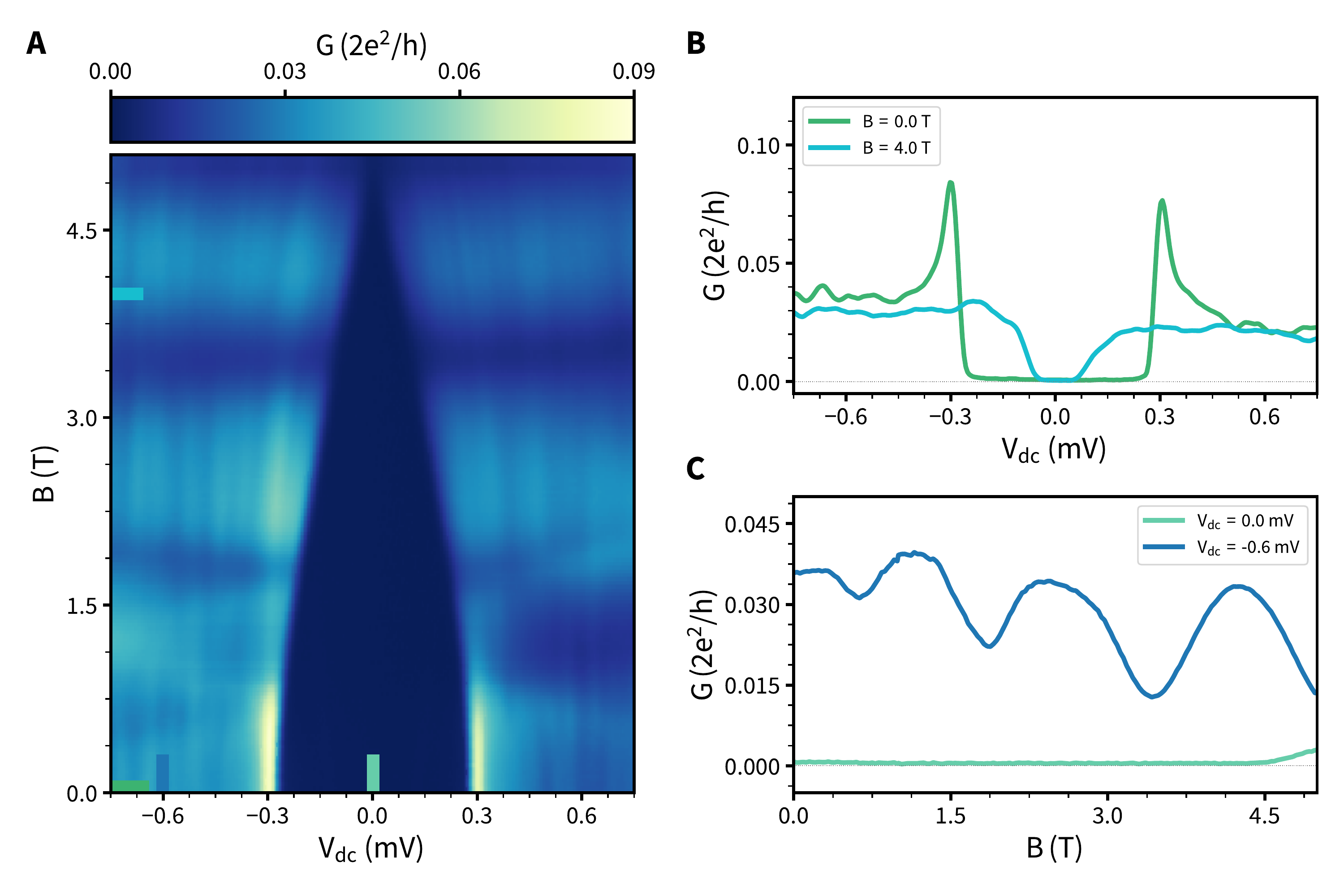}
	\caption{\textbf{Device I.} Example of a 2-facet device fabricated with the single-shot technique. Panel \textbf{(A)} presents differential conductance map as a function of magnetic field, with a hard gap up to $B$ = 4.5\,T. Panel \textbf{(B)} shows line cuts of the differential conductance at $B$ = 0\,T and $B$ = 4\,T. Panel \textbf{(C)} illustrates the evolution of the out-of-gap ($V_{\rm dc}$ = 0.6\,mV) and in-gap ($V_{\rm dc}$ = 0\,mV) conductance as a function of parallel magnetic field. Oscillatory behavior of the out-of-gap conductance indicates the presence of a quantum dot in the vicinity of the tunnel junction}
\label{fig:Fig_s7}
\end{figure}
\clearpage

\begin{figure}[ht]
	\includegraphics[width=\textwidth]{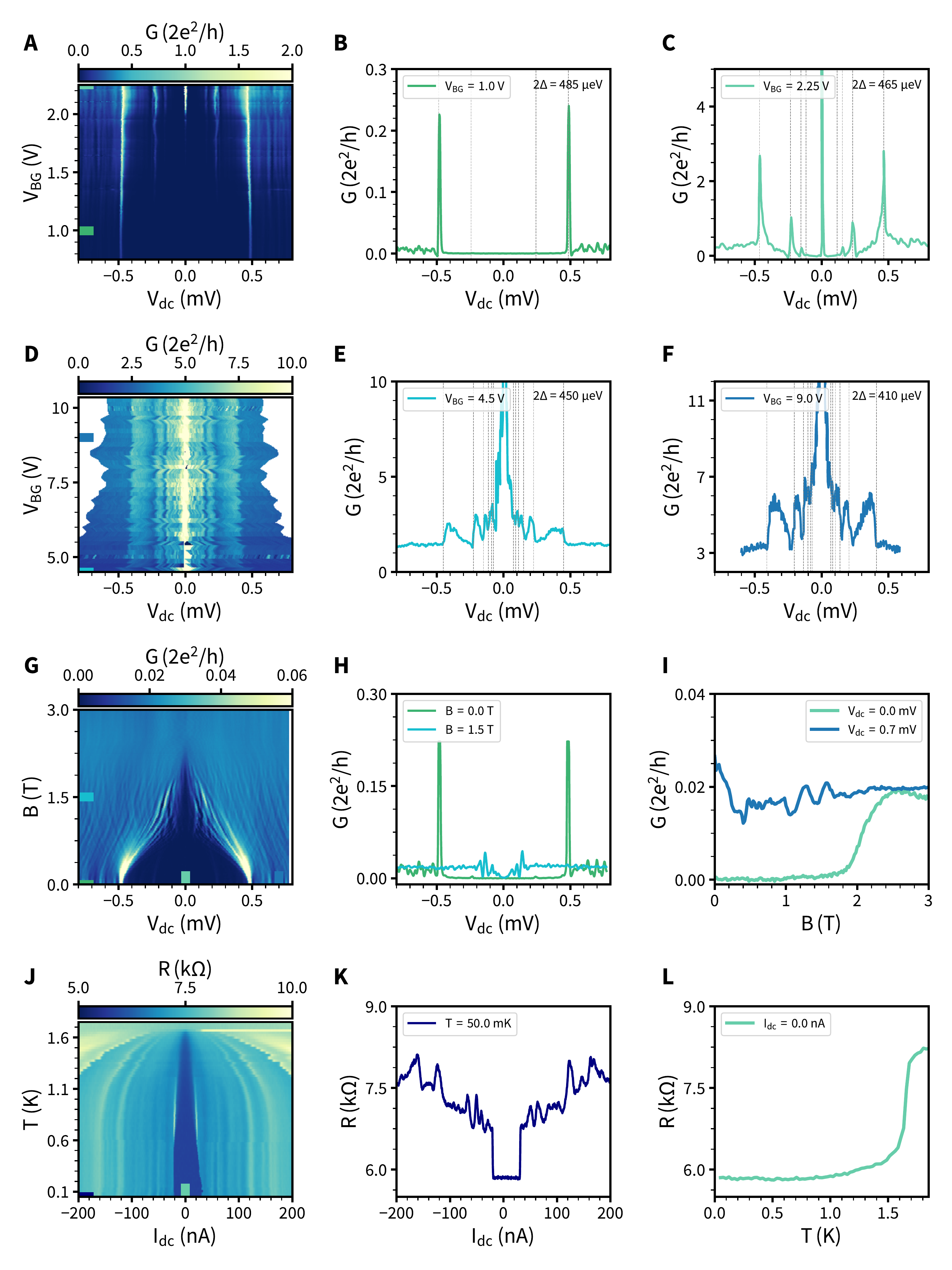}
\label{fig:Fig_s8}
\end{figure}
\clearpage

\captionof{figure}{\textbf{Device J.} Collection of data on a nanowire Josephson junction with Al/Pt. Panel \textbf{(A)} is a map of the differential conductance as a function of back-gate voltage in a low-transmission regime. Panels \textbf{(B)} and \textbf{(C)} are differential conductance traces as a function of voltage bias. Differential conductance at a higher transmission is presented on panel \textbf{(D)} with corresponding conductance traces shown in panels \textbf{(E,F)}. Panel \textbf{(G)} depicts the parallel magnetic field evolution of the differential conductance, with a critical field of $B_{\rm c}$ = 2\,T. Panel \textbf{(J)} shows the temperature evolution of the switching current, Panel \textbf{(K)} show the differential resistance trace at $T$ = 50\,mK and panel \textbf{(L)} illustrates the in-gap resistance, indicating a critical temperature of $T_{\rm c}$ = 1.6\,K. In comparison to InSb-Al Josephson junctions \cite{Heedt:2021_NC,Borsoi:2021_AFM}, the reduced $T_{\rm c}$ in combination with a modest increase in $B_{\rm c}$ indicates that the orbital contribution of the magnetic field dominates the behavior of devices with moderately thick (12.5\,nm) Al shells. About 5.8\,k$\Omega$ resistance in panels \textbf{J-L}, corresponds to the fridge line resistance.}

\begin{figure}[ht]
	\includegraphics[width=\textwidth]{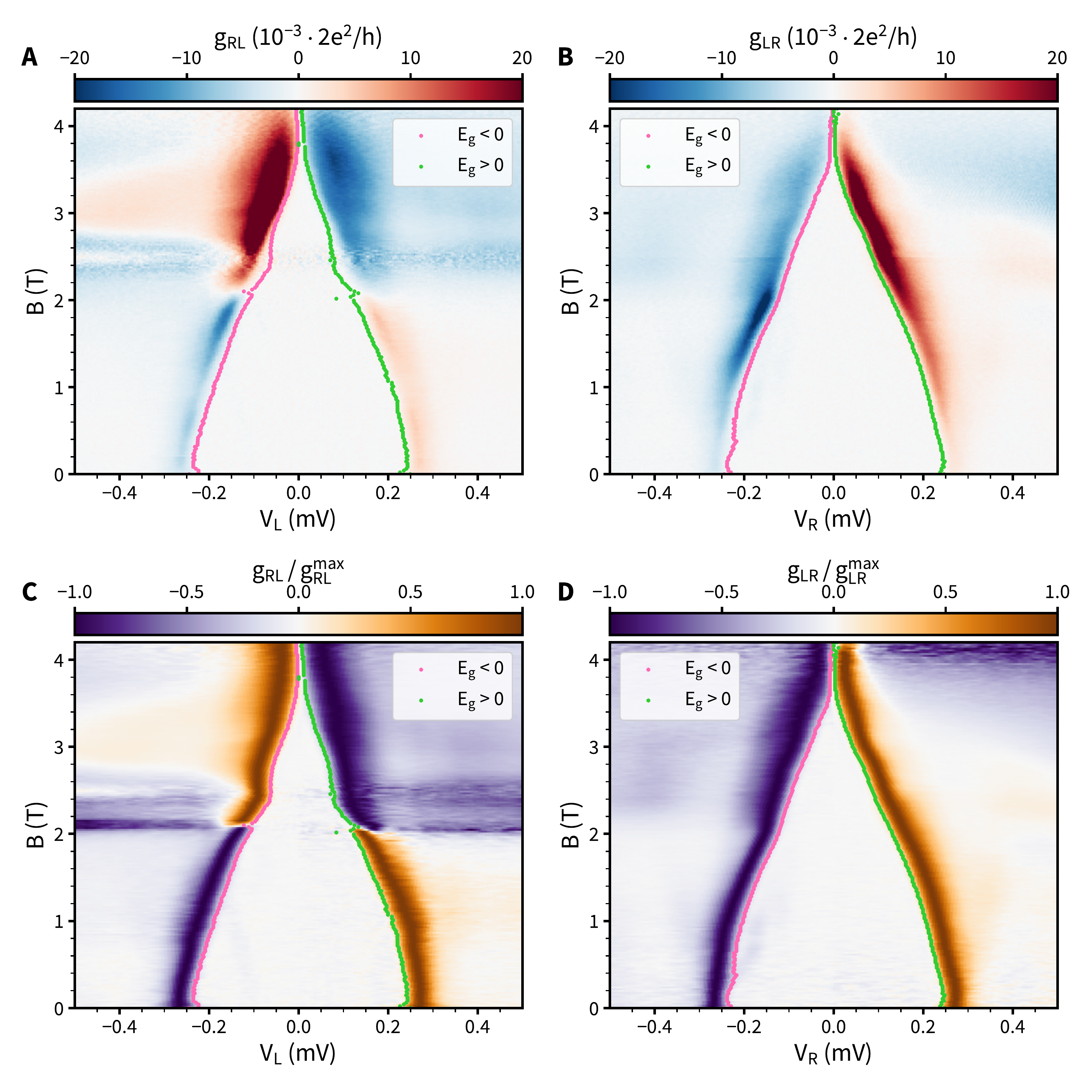}
	\caption{\textbf{Device C.} Extraction of the induced gap size from non-local signals $g_{\rm LR}$ (\textbf{A}) and $g_{\rm LR}$ (\textbf{B}). Negative energy values are shown in pink, positive energy values in green. In panels \textbf{C} and \textbf{D}, the extracted gap values are overlayed with the normalized and Savitzky-Golay filtered data.}
\label{fig:Fig_s9}
\end{figure}
\clearpage

\newpage
\bibliography{scibib2}

\bibliographystyle{Science}